\documentclass
[nobibnotes,reprint,aps,prl,twocolumn,superscriptaddress,showpacs,preprintnumbers,amsmath,amssymb,floatfix]{revtex4-1}

\usepackage{graphicx}
\usepackage[mathlines]{lineno}
\usepackage{xspace} 
\usepackage{wasysym}

\newcommand{\gev}{GeV/c$^2$\xspace}

\newcommand{\cf}{$^{252}$Cf\xspace}
\newcommand{\kevee}{keV$_{\mathrm{ee}}$\xspace}

\begin{document}
\title{Search for annual modulation in low-energy CDMS-II data}

\author{Z.~Ahmed} \affiliation{Division of Physics, Mathematics, and Astronomy, California Institute of Technology, Pasadena, CA 91125, USA} 
\author{D.S.~Akerib} \affiliation{Department of Physics, Case Western Reserve University, Cleveland, OH  44106, USA} 
\author{A.J.~Anderson} \affiliation{Department of Physics, Massachusetts Institute of Technology, Cambridge, MA 02139, USA} 
\author{S.~Arrenberg} \affiliation{Physics Institute, University of Z\"{u}rich, Winterthurerstr. 190, CH-8057, Switzerland}  
\author{C.N.~Bailey} \affiliation{Department of Physics, Case Western Reserve University, Cleveland, OH  44106, USA} 
\author{D.~Balakishiyeva} \affiliation{Department of Physics, University of Florida, Gainesville, FL 32611, USA} 
\author{L.~Baudis} \affiliation{Physics Institute, University of Z\"{u}rich, Winterthurerstr. 190, CH-8057, Switzerland}  
\author{D.A.~Bauer} \affiliation{Fermi National Accelerator Laboratory, Batavia, IL 60510, USA} 
\author{P.L.~Brink} \altaffiliation{Current Affiliation:  SLAC National Accelerator Laboratory/Kavli Institute for Particle Astrophysics and Cosmology, 2575 Sand Hill Road, Menlo Park 94025, CA} \affiliation{Department of Physics, Stanford University, Stanford, CA 94305, USA}
\author{T.~Bruch} \affiliation{Physics Institute, University of Z\"{u}rich, Winterthurerstr. 190, CH-8057, Switzerland} 
\author{R.~Bunker} \affiliation{Department of Physics, Syracuse University, Syracuse, NY 13244, USA} 
\author{B.~Cabrera} \affiliation{Department of Physics, Stanford University, Stanford, CA 94305, USA} 
\author{D.O.~Caldwell} \affiliation{Department of Physics, University of California, Santa Barbara, CA 93106, USA} 
\author{J.~Cooley} \affiliation{Department of Physics, Southern Methodist University, Dallas, TX 75275, USA} 
\author{P.~Cushman} \affiliation{School of Physics \& Astronomy, University of Minnesota, Minneapolis, MN 55455, USA} 
\author{M.~Daal} \affiliation{Department of Physics, University of California, Berkeley, CA 94720, USA} 
\author{F.~DeJongh} \affiliation{Fermi National Accelerator Laboratory, Batavia, IL 60510, USA} 
\author{P.C.F.~Di Stefano} \affiliation{Department of Physics, Queen's University, Kingston ON, Canada K7L 3N6} 
\author{M.R.~Dragowsky} \affiliation{Department of Physics, Case Western Reserve University, Cleveland, OH  44106, USA} 
\author{S.~Fallows} \affiliation{School of Physics \& Astronomy, University of Minnesota, Minneapolis, MN 55455, USA} 
\author{E.~Figueroa-Feliciano} \affiliation{Department of Physics, Massachusetts Institute of Technology, Cambridge, MA 02139, USA} 
\author{J.~Filippini} \affiliation{Division of Physics, Mathematics, and Astronomy, California Institute of Technology, Pasadena, CA 91125, USA} 
\author{J.~Fox} \affiliation{Department of Physics, Queen's University, Kingston ON, Canada K7L 3N6} 
\author{M.~Fritts} \affiliation{School of Physics \& Astronomy, University of Minnesota, Minneapolis, MN 55455, USA} 
\author{S.R.~Golwala} \affiliation{Division of Physics, Mathematics, and Astronomy, California Institute of Technology, Pasadena, CA 91125, USA} 
\author{J.~Hall} \affiliation{Fermi National Accelerator Laboratory, Batavia, IL 60510, USA} 
\author{S.A.~Hertel}  \email{Corresponding author: hertel@mit.edu} \affiliation{Department of Physics, Massachusetts Institute of Technology, Cambridge, MA 02139, USA}  
\author{T.~Hofer} \affiliation{School of Physics \& Astronomy, University of Minnesota, Minneapolis, MN 55455, USA} 
\author{D.~Holmgren} \affiliation{Fermi National Accelerator Laboratory, Batavia, IL 60510, USA} 
\author{L.~Hsu} \affiliation{Fermi National Accelerator Laboratory, Batavia, IL 60510, USA} 
\author{M.E.~Huber} \affiliation{Department of Physics, University of Colorado Denver, Denver, CO 80217, USA} 
\author{O.~Kamaev} \affiliation{Department of Physics, Queen's University, Kingston ON, Canada K7L 3N6} 
\author{M.~Kiveni} \affiliation{Department of Physics, Syracuse University, Syracuse, NY 13244, USA} 
\author{M.~Kos} \affiliation{Department of Physics, Syracuse University, Syracuse, NY 13244, USA} 
\author{S.W.~Leman} \affiliation{Department of Physics, Massachusetts Institute of Technology, Cambridge, MA 02139, USA} 
\author{S.~Liu} \affiliation{Department of Physics, Queen's University, Kingston ON, Canada K7L 3N6} 
\author{R.~Mahapatra} \affiliation{Department of Physics, Texas A\&M University, College Station, TX 77843, USA} 
\author{V.~Mandic} \affiliation{School of Physics \& Astronomy, University of Minnesota, Minneapolis, MN 55455, USA} 
\author{K.A.~McCarthy} \affiliation{Department of Physics, Massachusetts Institute of Technology, Cambridge, MA 02139, USA} 
\author{N.~Mirabolfathi} \affiliation{Department of Physics, University of California, Berkeley, CA 94720, USA} 
\author{D.C.~Moore} \affiliation{Division of Physics, Mathematics, and Astronomy, California Institute of Technology, Pasadena, CA 91125, USA} 
\author{H.~Nelson} \affiliation{Department of Physics, University of California, Santa Barbara, CA 93106, USA} 
\author{R.W.~Ogburn} \affiliation{Division of Physics, Mathematics, and Astronomy, California Institute of Technology, Pasadena, CA 91125, USA} 
\author{A.~Phipps}\affiliation{Department of Physics, University of California, Berkeley, CA 94720, USA}
\author{K.~Prasad} \affiliation{Department of Physics, Texas A\&M University, College Station, TX 77843, USA} 
\author{M.~Pyle} \affiliation{Department of Physics, Stanford University, Stanford, CA 94305, USA} 
\author{X.~Qiu} \affiliation{School of Physics \& Astronomy, University of Minnesota, Minneapolis, MN 55455, USA} 
\author{W.~Rau} \affiliation{Department of Physics, Queen's University, Kingston ON, Canada K7L 3N6} 
\author{A.~Reisetter} \affiliation{School of Physics \& Astronomy, University of Minnesota, Minneapolis, MN 55455, USA} 
\author{Y.~Ricci} \affiliation{Department of Physics, Queen's University, Kingston ON, Canada K7L 3N6} 
\author{T.~Saab} \affiliation{Department of Physics, University of Florida, Gainesville, FL 32611, USA} 
\author{B.~Sadoulet}\affiliation{Department of Physics, University of California, Berkeley, CA 94720, USA} \affiliation{Lawrence Berkeley National Laboratory, Berkeley, CA 94720, USA} 
\author{J.~Sander} \affiliation{Department of Physics, Texas A\&M University, College Station, TX 77843, USA} 
\author{R.W.~Schnee} \affiliation{Department of Physics, Syracuse University, Syracuse, NY 13244, USA} 
\author{D.~Seitz}\affiliation{Department of Physics, University of California, Berkeley, CA 94720, USA}
\author{B.~Serfass} \affiliation{Department of Physics, University of California, Berkeley, CA 94720, USA} 
\author{D.~Speller} \affiliation{Department of Physics, University of California, Berkeley, CA 94720, USA} 
\author{K.M.~Sundqvist} \affiliation{Department of Physics, University of California, Berkeley, CA 94720, USA}
\author{M.~Tarka} \affiliation{Physics Institute, University of Z\"{u}rich, Winterthurerstr. 190, CH-8057, Switzerland} 
\author{R.B.~Thakur } \affiliation{Fermi National Accelerator Laboratory, Batavia, IL 60510, USA} 
\author{A.N.~Villano} \affiliation{School of Physics \& Astronomy, University of Minnesota, Minneapolis, MN 55455, USA} 
\author{B.~Welliver} \affiliation{Department of Physics, University of Florida, Gainesville, FL 32611, USA} 
\author{S.~Yellin} \affiliation{Department of Physics, Stanford University, Stanford, CA 94305, USA} 
\author{J.~Yoo} \affiliation{Fermi National Accelerator Laboratory, Batavia, IL 60510, USA} 
\author{B.A.~Young} \affiliation{Department of Physics, Santa Clara University, Santa Clara, CA 95053, USA} 
\author{J.~Zhang} \affiliation{School of Physics \& Astronomy, University of Minnesota, Minneapolis, MN 55455, USA} 
\collaboration{The CDMS II Collaboration}

\noaffiliation

\begin{abstract}
We report limits on annual modulation of the low-energy event rate from the Cryogenic Dark Matter Search (CDMS~II) experiment at the Soudan Underground Laboratory.  Such a modulation could be produced by interactions from Weakly Interacting Massive Particles (WIMPs) with masses $\sim$10~\gev.  We find no evidence for annual modulation in the event rate of 
veto-anticoincident single-detector interactions consistent with nuclear recoils, and constrain the magnitude of any modulation to $<$0.06\,event\,[keV$_{\mathrm{nr}}$\,kg\,day]$^{-1}$ in the 5\textendash11.9\,keV$_{\mathrm{nr}}$ energy range at the 99\% confidence level.  These results disfavor an explanation for the reported modulation in the 1.2\textendash3.2\,keV$_{\mathrm{ee}}$ energy range in CoGeNT in terms of nuclear recoils resulting from elastic scattering of WIMPs at $>$98\% confidence.  For events consistent with electron recoils, no significant modulation is observed for either single- or multiple-detector interactions in the 3.0\textendash7.4\,keV$_{\mathrm{ee}}$ range.
\end{abstract}


\maketitle


Astrophysical observations indicate that the vast majority of matter in the universe consists of non-baryonic, non-luminous dark matter~\cite{Bertone:2004pz,Gaitskell:2004gd}.  Weakly Interacting Massive Particles (WIMPs) are leading candidates for this dark matter since they would be thermally produced in the early universe in the correct abundance to account for the observed relic density~\cite{Bertone:2004pz,*Gaitskell:2004gd,Jungman:1995df}.  Possible experimental signals from DAMA/LIBRA~\cite{Bernabei:2008yi,*Bernabei:2010}, CoGeNT~\cite{Aalseth:2010vx,*Aalseth:2011ce} and CRESST-II~\cite{Angloher:2011} have led to significant recent interest in WIMPs with masses $\sim$10~\gev and spin-independent WIMP-nucleon cross sections $\sim$10$^{-41}$\textendash10$^{-40}$\,cm$^2$ (\emph{e.g.}~\cite{Hooper:2012fk,Fox:2011kx,Kopp:2011uq,*Belli:2011fa}).  
 
The WIMP scattering rate is expected to annually modulate due to the relative motion of the earth through the local dark-matter halo~\cite{Drukier:1986}.  The presence of an annually modulating component in the observed interaction rate can identify a WIMP signal in the presence of significant unmodulated backgrounds.  This modulation signature is especially useful for WIMPs with masses $\sim$10\,\gev which would primarily produce recoils with energies just above the detection threshold, where the rejection of backgrounds that can mimic a WIMP signal is less powerful.   Both the DAMA/LIBRA~\cite{Bernabei:2008yi,*Bernabei:2010} and CoGeNT~\cite{Aalseth:2010vx,*Aalseth:2011ce} experiments claim evidence for such a modulating signal in their data.


The CDMS~II experiment consists of an array of cryogenic germanium and silicon detectors which measure both the ionization and athermal phonon energy deposited by each particle interaction~\cite{Akerib:2005zy,Ahmed:2008eu,*CDMSScience:2010}.  The ratio of the ionization to phonon energy allows discrimination of the expected WIMP nuclear-recoil signal from more prevalent electron-recoil backgrounds on an event-by-event basis.  At the $\lesssim$10\,keV recoil energies expected from WIMPs with masses $\lesssim$10\,\gev, electron-recoil rejection is less effective, because the ionization signal is comparable to readout noise.  Even with this limited rejection, the observed interaction rate in the CDMS~II germanium detectors has been shown in a previous analysis~\cite{Ahmed:2011le} to disfavor an explanation for the DAMA/LIBRA and CoGeNT signals in terms of spin-independent elastic scattering of low-mass WIMPs.  

If, as recently suggested, only a small fraction of the low-energy excess events in CoGeNT are due to WIMPs, then constraints from CDMS~II may be avoided~\cite{Hooper:2012fk,Collar:TAUP}. In this case, if CoGeNT's annual modulation is due to WIMPs, the fractional variation is several times larger than expected for a ``standard'' halo with a Maxwellian velocity distribution~\cite{Hooper:2012fk,Kelso:2011vn}.  In addition, the energy spectrum of the modulation extends to higher energies than expected for a standard halo~\cite{Fox:2011kx,Kelso:2011vn,Schwetz:2011,Farina:2011}. Such large modulation fractions and hard spectra might be possible if the halo exhibits (non-standard) local substructure~\cite{Hooper:2012fk,Kelso:2011vn,Fox:2011kx}.  To test such a scenario, this Letter searches for a corresponding annual modulation in the CDMS~II germanium data.  This analysis does not cover the full energy range of the CoGeNT modulation, restricting itself to energies above 5\,keV$_{\mathrm{nr}}$ (which, due to quenching~\cite{Lindhard:1963,Lewin:1995rx}, corresponds to 1.2\,keV$_{\mathrm{ee}}$ in the standard CoGeNT energy scale~\cite{Collar:priv}).  Because germanium serves as the target material for both CDMS~II and CoGeNT, these results provide a check of whether the reported modulating signal is due to WIMPs that is less model-dependent than recent results from XENON10~\cite{Angle:2011} and XENON100~\cite{Aprile:2011ba}.

The data analyzed here were collected over nearly two annual cycles, from October 2006 to September 2008, using all 30 Z-sensitive Ionization and Phonon (ZIP) detectors installed at the Soudan Underground Laboratory~\cite{Akerib:2005zy,Ahmed:2008eu,*CDMSScience:2010}.  Data-quality and detector selection criteria are identical to the previous analysis of the low-energy CDMS~II nuclear-recoil spectrum in~\cite{Ahmed:2011le}. Only the 8 germanium detectors with the lowest trigger thresholds were used to search for WIMP interactions, while all 30 detectors were used to veto events with interactions in multiple detectors. 

Following~\cite{Ahmed:2011le},  the nuclear-recoil energy scale is based on the phonon measurement, which is corrected by $\sim$20\% to take into account the fraction of the total phonon signal arising from the Neganov-Luke phonons~\cite{Neganov:1985,*Luke:1988} generated by the charge-carrier drift across the detectors. The Neganov-Luke phonon contribution for nuclear recoils in \cf calibration data was directly measured for the recoil-energy range of this analysis. As in~\cite{Ahmed:2011le}, the phonon energy scale for electron recoils was conservatively calibrated, ensuring to the 90\% C.L. that the 1.3 and 10.4\,keV activation-line energies were not underestimated. 

The maximum energy considered in this analysis was 11.9\,keV$_{\mathrm{nr}}$, matching the highest energy observable by the CoGeNT ``LG'' (low gain) channel~\cite{Collar:priv}.  Because time-dependent variations in the CDMS trigger thresholds could mimic or hide a modulation in the event rate, the energy threshold for this analysis was conservatively chosen to be 5\,keV$_{\mathrm{nr}}$, high enough that events are triggered with essentially perfect efficiency.  To avoid bias, trigger efficiency was measured throughout the exposure, using events for which at least one other detector triggered.  For the 5\textendash11.9 keV$_{\mathrm{nr}}$ energy range, combining all detectors and all time bins yielded 4350 events in this unbiased sample, only 3 of which failed to trigger.  These missed triggers were each in a different detector and were uniformly spaced throughout the considered energy range.

Because CDMS~II uses a phonon-based energy scale (at these low energies), and CoGeNT uses an ionization-based energy scale, quenching causes the two experiments to exhibit different mappings between energies assuming nuclear recoils (such as the energy range of this analysis, 5.0\textendash11.9\,keV$_{\mathrm{nr}}$) and energies assuming electron recoils.  For electron recoils with the same total phonon signal in the CDMS~II experiment, the equivalent recoil-energy interval is 3.0\textendash7.4\,keV$_{\mathrm{ee}}$, due to the larger Neganov-Luke phonon contribution. Analogously, for electron recoils with the same total \emph{ionization} signal in the CoGeNT experiment, the equivalent recoil-energy interval is 1.2\textendash3.2\,keV$_{\mathrm{ee}}$, where we apply CoGeNT's measured ionization yield for nuclear recoils~\cite{Collar:priv,Barbeau:thesis}.

Detector stability was monitored throughout data taking with quality cuts, removing periods of abnormal detector performance~\cite{Ahmed:2008eu,*CDMSScience:2010}.  For consistency with previous work, we followed~\cite{Ahmed:2011le} and removed data taken during the 20 days following exposure of the detectors to a neutron calibration source. After removing these time periods, a total of 241\,kg\,days raw exposure were considered, as in~\cite{Ahmed:2011le}. To allow checks for stability to be applied to multiple-scatter events, an additional cut was introduced to eliminate electronics ``glitch'' events, for which phonon pulses were detected above threshold in more than 15 detectors simultaneously.

Events inconsistent with WIMP interactions were rejected. Since modulation of data-selection cut efficiencies could mimic or hide a modulation in the event rate, selection criteria were designed to have constant acceptance with time, and any residual modulation in the cut efficiencies was constrained using events sampled throughout the data taking period.  Since WIMPs have a negligibly small probability of interacting more than once in the apparatus, events with energy deposited in more than a single detector (``singles cut") or in the active scintillator veto (``veto cut'') were removed.  The glitch cut, veto cut and singles cut have a combined efficiency $>$97\%, with negligible time-dependent variation.  

Events were further required to have ionization signals consistent with noise in the outer charge electrode of the detector (``$Q$-inner cut"). To search for a nuclear-recoil signal of WIMP origin, the ionization energy was required to be within $\pm$2$\sigma$ of the mean of the energy-dependent nuclear-recoil distribution from calibration data (``nuclear-recoil cut") in the main analysis described below.  This ionization-based selection increases the sensitivity of this analysis to a modulating signal relative to the more restrictive ionization-based selection used in~\cite{Ahmed:2011le}, provided that backgrounds do not modulate.  To explore different physics or instrumental origins of a potential signal, we also applied our modulation analysis to two additional event samples consisting of either single-scatter or multiple-scatter events with no ionization-based nuclear-recoil cut. The quality, glitch, veto and $Q$-inner cuts were always maintained.

These various cuts result in an efficiency $\varepsilon \left( {t,E,d} \right)$ that depends on the time $t$, the deposited energy $E$ and the detector $d$.  With notations following the descriptions above, the total efficiency for our primary ``WIMP-candidate'' sample can be written as 
\begin{align}
\label{eqs:efficiency}
   \varepsilon \left( {t,E,d} \right) = &  \varepsilon _{{\rm{glitch}}} \varepsilon_{{\rm{trigger}}} \varepsilon_{{\rm{singles}}} \left( {d} \right) \varepsilon_{{\rm{veto}}} \times \\
                                                             &  \varepsilon_{{\rm{Qinner}}} \left( {t,E,d} \right)\varepsilon _{{\rm{NuclearRecoil}}} \left( {t,E,d} \right) ,\nonumber 
\end{align} 
where we have explicitly identified the dependence on time $t$, energy $E$ and detector $d$ for each of the cuts. For the event samples that remove the nuclear-recoil and singles cuts, the corresponding efficiencies, $\varepsilon _{{\rm{singles}}}$ and $\varepsilon _{{\rm{NuclearRecoil}}}$, are ignored.

In the lower portion (5\textendash7.3 keV$_{\mathrm{nr}}$) of the energy range considered, the rate of nuclear-recoil candidate events measured in this analysis is 0.28$\pm$0.03\,[keV$_{\mathrm{nr}}$~kg~day]$^{-1}$, while the maximum-likelihood estimate for the CoGeNT modulation amplitude is 0.35\,[keV$_{\mathrm{nr}}$~kg~day]$^{-1}$.  The corresponding numbers for the entire energy interval considered are 0.15$\pm$0.01\,[keV$_{\mathrm{nr}}$~kg~day]$^{-1}$ for CDMS and 0.16\,[keV$_{\mathrm{nr}}$~kg~day]$^{-1}$ for CoGeNT.  In both cases, a modulation of the magnitude observed by CoGeNT would require a modulation fraction in CDMS of $\sim$100\%. 

We test whether the $Q$-inner and nuclear-recoil cut efficiencies are sufficiently constant using calibration data collected throughout the time period used in the analysis. For each cut, for a given time interval $\gamma$, and detector $d$, we measure $P_{\gamma d}$ events passing the cut and $F_{\gamma d}$ failing. Note that the time intervals of the efficiency data are not coincident with the low-background time intervals, but are suitably distributed over the whole data-taking period.  We then maximize the likelihood appropriate for a binomial distribution
\begin{equation}
l  = \prod\limits_{\gamma d} {\varepsilon _{\gamma d}^{P_{\gamma d} } \left( {1 - \varepsilon _{\gamma d}^{} } \right)^{F_{\gamma d} } } .
\label{eqs:likelihood_cut}
\end{equation}
The efficiencies ${\varepsilon _{\gamma d}}$ are written as 
\begin{equation}
\varepsilon _{\gamma d}^{}  = \varepsilon _{d} \left\{ {1 + A \cos \left[ {\omega \left( {t_\gamma   - \phi } \right) }\right]} \right\} ,
\label{eqs:efficiency_cut}
\end{equation}
where $\omega  = 2\pi /365.24$\,day$^{-1}$.  For the chosen cut and energy interval we fit for the detector-dependent unmodulated efficiency $\varepsilon _{d}$, the detector-independent relative modulation amplitude $A$, and the phase $\phi$ (measured from Jan. 1st), while requiring the efficiency to be $\leq$1.  We generate $10^4$ artificial realizations of the model under consideration, and determine the confidence regions for the modulation and phase using the Feldman-Cousins method~\cite{Feldman:1998}.  This analysis indicates that the maximum efficiency modulation allowed by our experimental measurement of the nuclear-recoil cut efficiency is 1.2\% at the 90\% C.L. In the case of the $Q$-inner cut, this upper limit is 2.3\%. 

In order to estimate the modulation in the observed event rate, we bin events into 16 time intervals, labeled by $\beta$, of $\sim$25 days each. We denote the center of each time bin as $t_\beta$ and its width as $\Delta t_\beta $. The number of events observed in the time interval $\beta$ in detector $d$ is $n_{\beta d}$. We construct a likelihood using the expected Poisson distribution for the $n_{\beta d}$
\begin{equation}
\ell   = \prod\limits_{\beta ,d} {e^{ - \mu _{\beta d} } \left( {\mu _{\beta d} } \right)^{n_{\beta d} } } ,
\label{eqs:likelihood}
\end{equation}
where factorial terms have been omitted for convenience.  In this equation, $\mu _{\beta d}$ is the expected number of events
\begin{equation}
\mu _{\beta d}  = \left\{ {\Gamma _{d}  + M  \cos \left[ { \omega \left( { t_\beta   - \phi} \right) } \right]}\right\} m_d \varepsilon _{\beta d} {f_{\beta d} } \Delta t_\beta   \Delta E  ,
\label{eqs:expected}
\end{equation}
where $\Gamma_d$ is the unmodulated rate in detector $d$, $M$ is the modulation amplitude, $m_d$ is the mass of detector $d$, $\varepsilon _{\beta d}$ is the appropriate efficiency using Eq.~\ref{eqs:efficiency}, ${f_{\beta d}}$ is the live-time fraction appropriate for the detector and time interval, $\Delta t_\beta$ is the time interval width, and $\Delta E$ is the energy interval width. 

Since the efficiency modulation allowed by the fits to Eq.~\ref{eqs:efficiency_cut} is much smaller than the physics effect we are testing, we need not add an additional term $\ell _{{\rm{eff}}} \left( {\varepsilon_{\beta d} } \right) $ in the likelihood, which would take into account such uncertainties in Eq.~\ref{eqs:efficiency_cut}.

\begin{figure}[tbp]
\begin{center}
\includegraphics[width=3.5in]{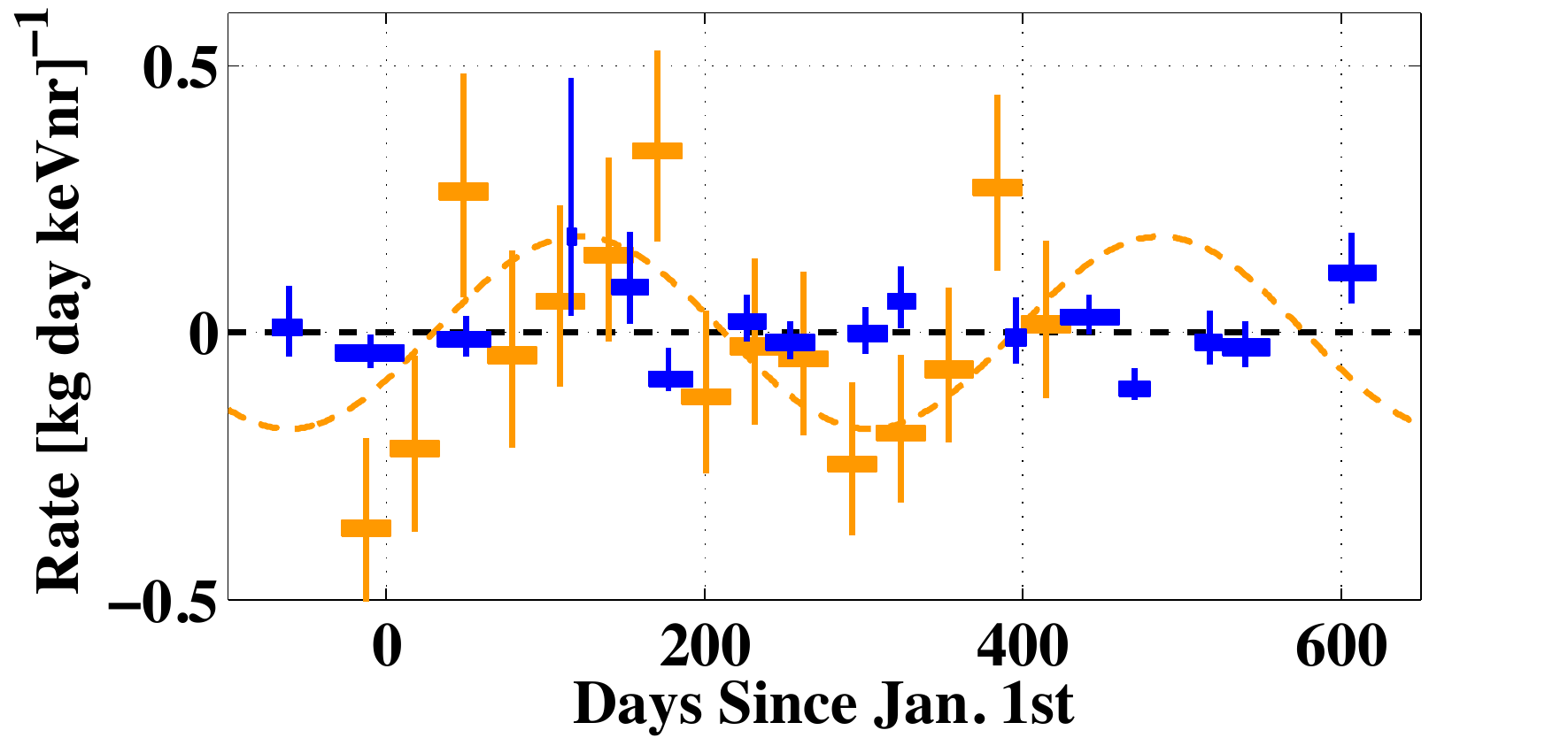}
\caption{\small (color online)  The rate of CDMS~II nuclear-recoil band events is shown for the 5.0\textendash11.9 keV$_{\mathrm{nr}}$ interval (dark blue), after subtracting the best-fit unmodulated rate, $\Gamma_d$, for each detector.  The horizontal bars represent the time bin extents, the vertical bars show $\pm1\sigma$ statistical uncertainties (note that one CDMS~II time bin is of extremely short duration).  The CoGeNT rates (assuming a nuclear-recoil energy scale) and maximum-likelihood modulation model in this energy range (light orange) are shown for comparison.  The CDMS exposure starts in late 2007, while the CoGeNT exposure starts in late 2009.}
\label{fig:RatevsTime}
\end{center}
\end{figure}

\begin{figure}[tbp]
\begin{center}
\includegraphics[width=3.5in]{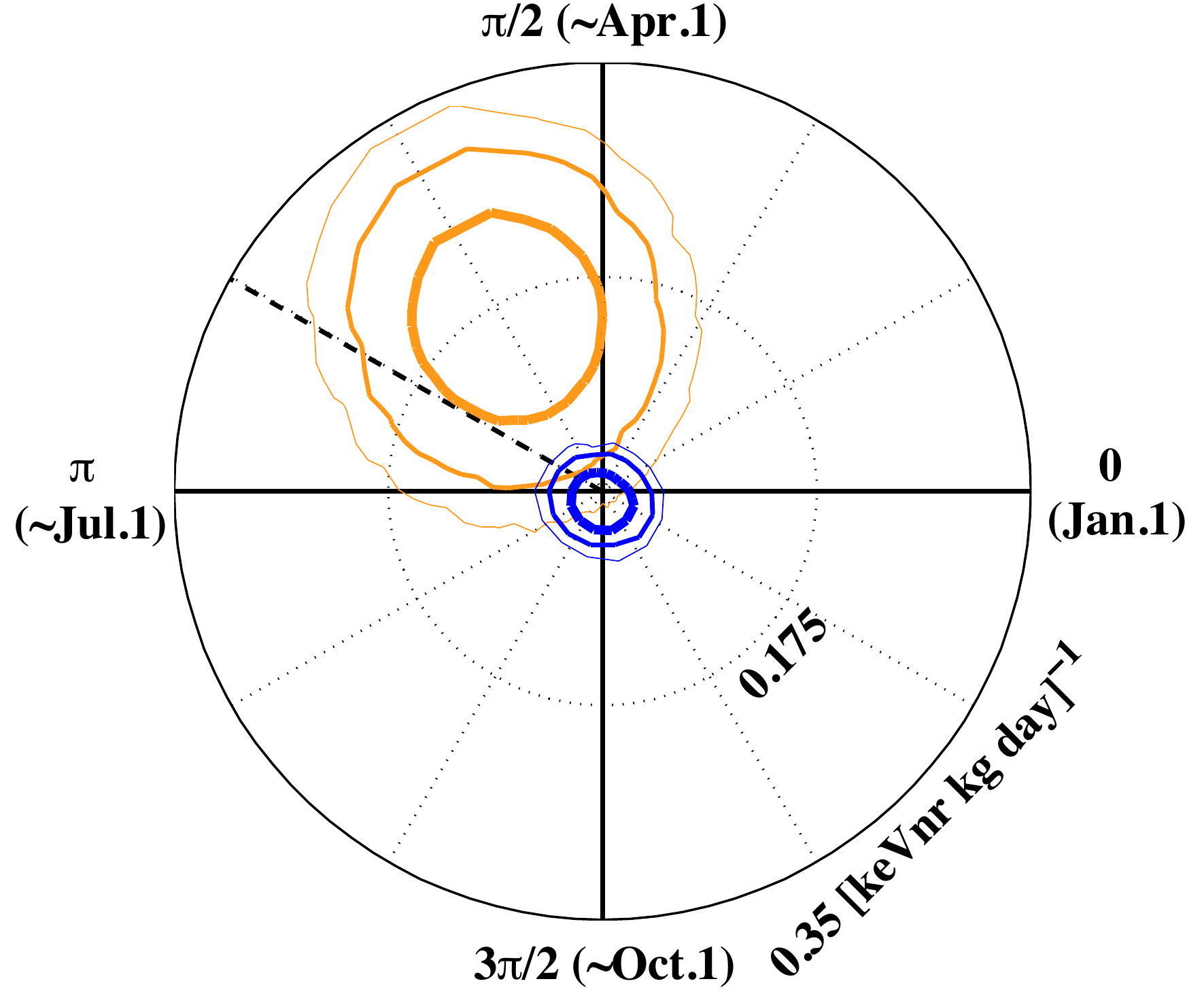}
\caption{\small (color online)  Allowed regions for annual modulation of CoGeNT (light orange) and the CDMS II nuclear-recoil sample (dark blue), for the 5.0\textendash11.9 keV$_{\mathrm{nr}}$ interval. In this and the following polar plot, a phase of 0 corresponds to January 1st, the phase of a modulation signal predicted by generic halo models (152.5 days) is highlighted by a dashed line, and 68\% (thickest), 95\%, and 99\% (thinnest) C.L. contours are shown.}
\label{fig:WIMPModulation}
\end{center}
\end{figure}

\begin{figure}[htbp]
\begin{center}
\includegraphics[width=3.5in]{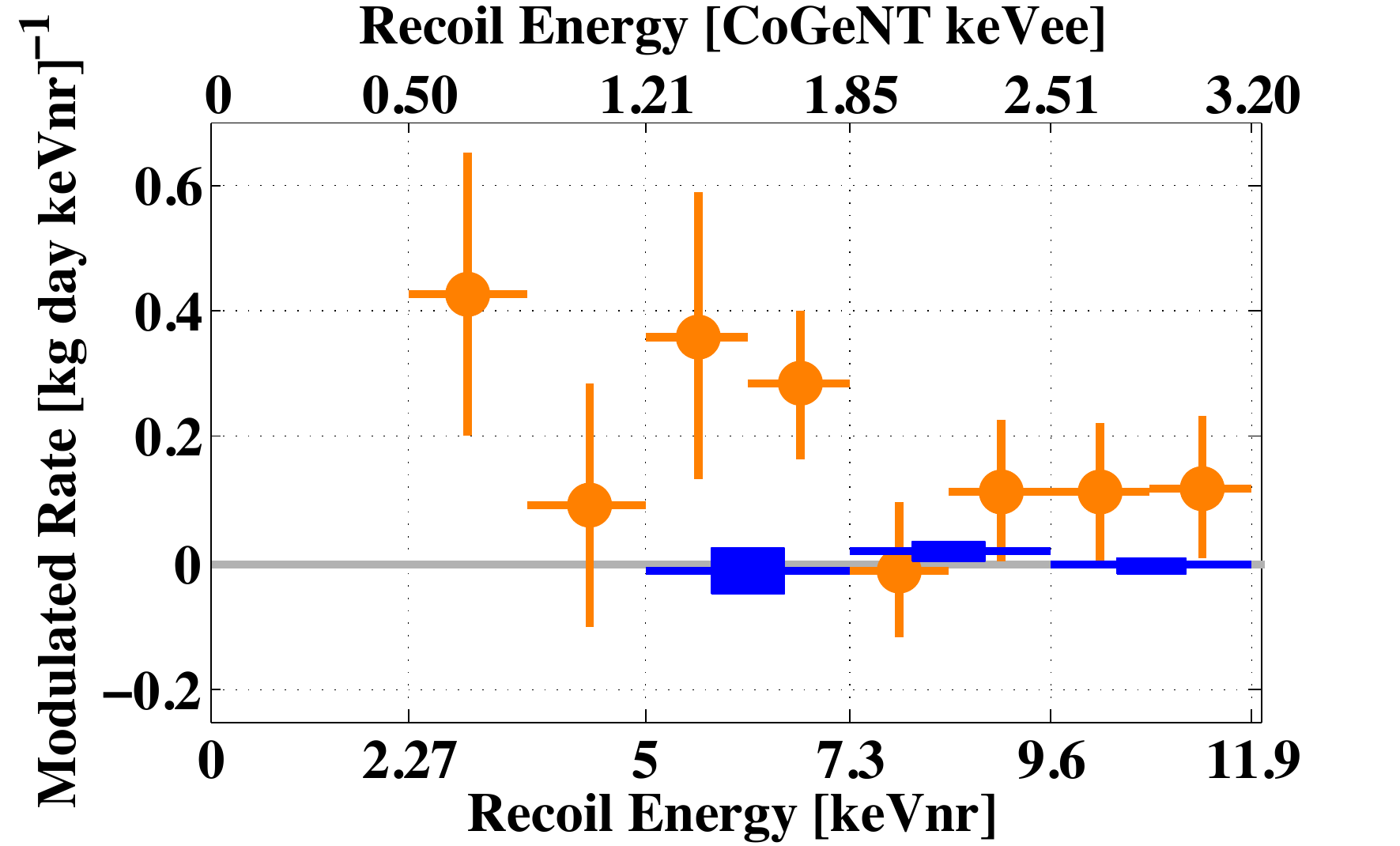}
\caption{\small (color online) Amplitude of modulation vs. energy, showing maximum-likelihood fits for both CoGeNT (light orange circles, 68\% confidence interval shown with vertical line) and CDMS nuclear-recoil singles (dark blue rectangles, 68\% confidence interval given by rectangle height).  The phase that best fits CoGeNT over all energies (106 days) was chosen for this representation.  The upper horizontal scale shows the electron-recoil-equivalent energy scale for CoGeNT events.  The 5\textendash11.9\,keV$_{\mathrm{nr}}$ energy range over which this analysis overlaps with the low-energy channel of CoGeNT has been divided into 3 (CDMS) and 6 (CoGeNT) equal-sized bins.}
\label{fig:NR_rmodVenergy_small}
\end{center}
\end{figure}

Figure~\ref{fig:RatevsTime} shows residual rates for WIMP candidate events, after subtracting the best-fit constant rates $\Gamma_d$ (found with modulated rate $M$ fixed at 0).  Using a Feldman-Cousins approach, we test modulation models [$M$, $\phi$] on the WIMP candidates, which consist of all events satisfying the data-selection cuts described above.  Figure~\ref{fig:WIMPModulation} shows that our observed WIMP-candidate event rate is consistent with a constant value.  All modulated rates in this energy range with amplitudes greater than 0.06 [keV$_{\mathrm{nr}}$ kg day]$^{-1}$ are excluded at the 99\% C.L.

For comparison, a similar analysis was carried out using the publicly available CoGeNT data~\cite{Collar:priv}.  Our analysis of CoGeNT data is consistent with previously published analyses~\cite{Hooper:2012fk,Fox:2011kx,Kelso:2011vn}.  Figure~\ref{fig:NR_rmodVenergy_small} shows the modulated spectrum of both CDMS~II and CoGeNT, assuming the phase (106 days) which best fits the CoGeNT data over the full CoGeNT energy range.  Compatibility between the annual modulation signal of CoGeNT and the absence of a significant signal in CDMS is determined by a likelihood-ratio test, which involves calculating $\lambda \equiv {\cal L}_{0}/{\cal L}_{1}$, where ${\cal L}_{0}$ is the combined maximum likelihood of the CoGeNT and CDMS data assuming both arise from the same simultaneous best-fit values of $M$ and $\phi$, while ${\cal L}_1$ is the product of the maximum likelihoods when the best-fit values are determined for each dataset individually.  The probability distribution function of $-2 \ln \lambda$ was mapped using simulation, and agreed with the $\chi^2$ distribution with two degrees of freedom, as expected in the asymptotic limit of large statistics and away from physical boundaries.  The simulation found only 82 of the 5$\times$10$^3$ trials had a likelihood ratio more extreme than was observed for the two experiments, confirming the asymptotic limit computation which indicated 98.3\% C.L. incompatibility between the annual-modulation signals of CoGeNT and CDMS for the 5.0\textendash11.9 keV$_{\mathrm{nr}}$ interval.

\begin{figure}[tbp]
\begin{center}
\includegraphics[width=3.3in]{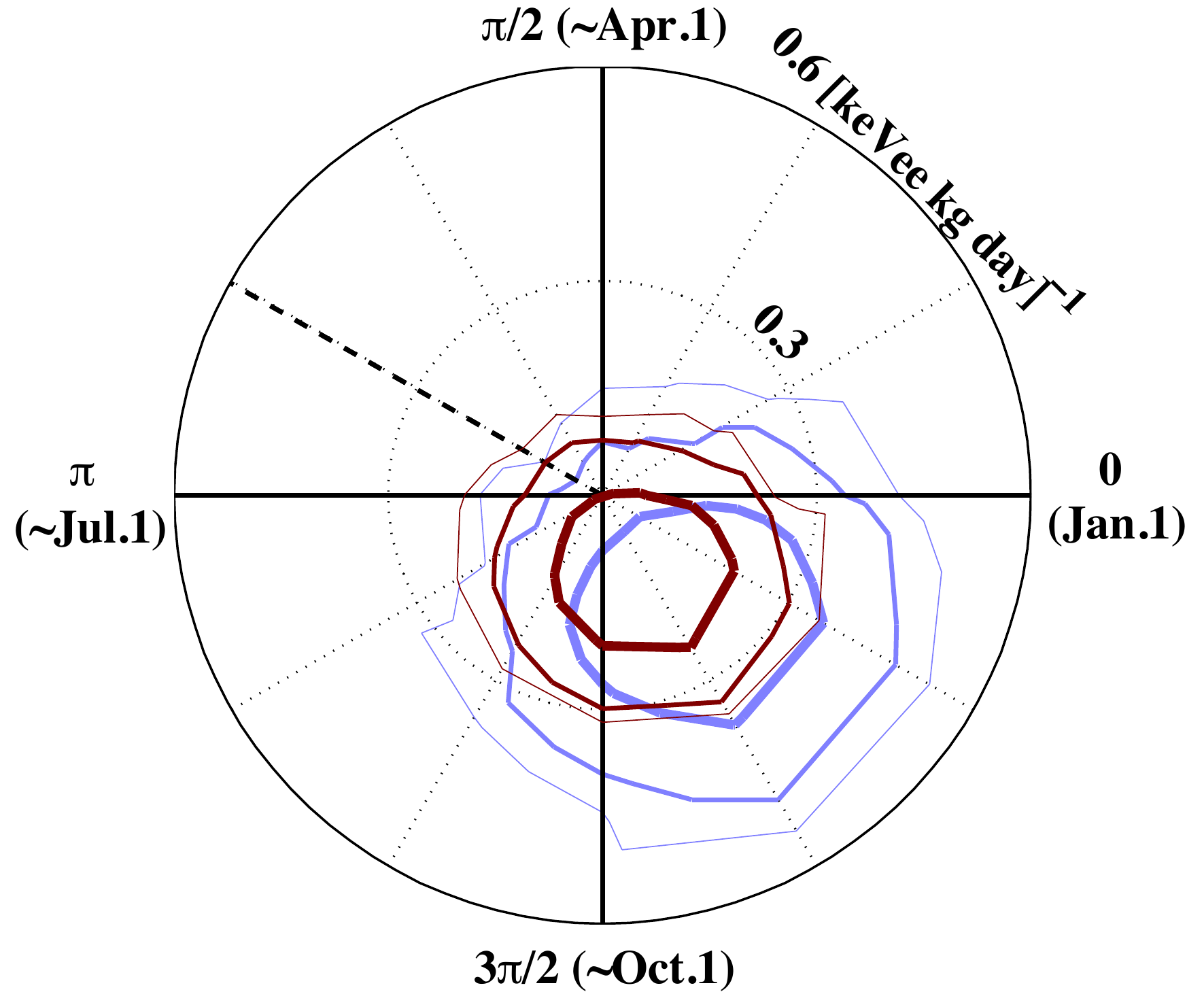}
\caption{\small (color online) Confidence limits on the amplitude and phase of annual modulation for two electron-recoil-dominated data samples: multiple scatters (light blue) and single scatters (dark red), as defined in the text for the interval 3.0\textendash7.4 keV$_{\mathrm{ee}}$.  These events are of the same total phonon energy (recoil + Neganov-Luke) as the nuclear-recoil band events of the main modulation analysis shown in Fig.~\ref{fig:WIMPModulation}, of 5.0\textendash11.9 keV$_{\mathrm{nr}}$.}
\label{fig:nonNRrates}
\end{center}
\end{figure}

We extend this analysis by applying the same method to CDMS II single-scatter and multiple-scatter events without applying the ionization-based nuclear-recoil cut. These samples are both dominated by electron recoils. Figure~\ref{fig:nonNRrates} shows the confidence intervals for the allowed modulation amplitudes and phases for these two samples, both of which are consistent with no modulation. For the energy range chosen for this analysis, there is not significant overlap with the corresponding CoGeNT energy range under the hypothesis of an electron-recoil modulation. Our minimum electron-equivalent energy is 3\,keV$_{\mathrm{ee}}$ compared to a 3.2\,keV$_{\mathrm{ee}}$ maximum energy for the CoGeNT low-energy channel. Consequently, this analysis cannot exclude the possibility of the modulation observed by CoGeNT being the result of electron recoils. The absence of modulation in the single-scatter and multiple-scatter events indicates the absence of strong systematic effects in our data. 

This analysis would not have been possible at this level of detail, without the willingness of the CoGeNT collaboration to provide public access to its data. We would also like to thank Neal Weiner for originally suggesting this analysis.  The CDMS collaboration gratefully acknowledges the contributions of numerous engineers and technicians; we would like to especially thank Jim Beaty, Bruce Hines, Larry Novak, Richard Schmitt and Astrid Tomada. In addition, we gratefully acknowledge assistance from the staff of the Soudan Underground Laboratory and the Minnesota Department of Natural Resources. This work is supported in part by the 
National Science Foundation (Grant Nos.\ AST-9978911, PHY-0542066, 
PHY-0503729, PHY-0503629, PHY-0503641, PHY-0504224, PHY-0705052, PHY-0801708, 
PHY-0801712, PHY-0802575, PHY-0847342, and PHY-0855525), by
the Department of Energy (Contracts DE-AC03-76SF00098, DE-FG02-91ER40688, 
DE-FG02-92ER40701, DE-FG03-90ER40569, and DE-FG03-91ER40618), by the Swiss National 
Foundation (SNF Grant No. 20-118119), and by NSERC Canada (Grant SAPIN 341314-07).

\bibliography{c38_AnnMod}

\begin{thebibliography}{35}%
\makeatletter
\providecommand \@ifxundefined [1]{%
 \@ifx{#1\undefined}
}%
\providecommand \@ifnum [1]{%
 \ifnum #1\expandafter \@firstoftwo
 \else \expandafter \@secondoftwo
 \fi
}%
\providecommand \@ifx [1]{%
 \ifx #1\expandafter \@firstoftwo
 \else \expandafter \@secondoftwo
 \fi
}%
\providecommand \natexlab [1]{#1}%
\providecommand \enquote  [1]{``#1''}%
\providecommand \bibnamefont  [1]{#1}%
\providecommand \bibfnamefont [1]{#1}%
\providecommand \citenamefont [1]{#1}%
\providecommand \href@noop [0]{\@secondoftwo}%
\providecommand \href [0]{\begingroup \@sanitize@url \@href}%
\providecommand \@href[1]{\@@startlink{#1}\@@href}%
\providecommand \@@href[1]{\endgroup#1\@@endlink}%
\providecommand \@sanitize@url [0]{\catcode `\\12\catcode `\$12\catcode
  `\&12\catcode `\#12\catcode `\^12\catcode `\_12\catcode `\%12\relax}%
\providecommand \@@startlink[1]{}%
\providecommand \@@endlink[0]{}%
\providecommand \url  [0]{\begingroup\@sanitize@url \@url }%
\providecommand \@url [1]{\endgroup\@href {#1}{\urlprefix }}%
\providecommand \urlprefix  [0]{URL }%
\providecommand \Eprint [0]{\href }%
\providecommand \doibase [0]{http://dx.doi.org/}%
\providecommand \selectlanguage [0]{\@gobble}%
\providecommand \bibinfo  [0]{\@secondoftwo}%
\providecommand \bibfield  [0]{\@secondoftwo}%
\providecommand \translation [1]{[#1]}%
\providecommand \BibitemOpen [0]{}%
\providecommand \bibitemStop [0]{}%
\providecommand \bibitemNoStop [0]{.\EOS\space}%
\providecommand \EOS [0]{\spacefactor3000\relax}%
\providecommand \BibitemShut  [1]{\csname bibitem#1\endcsname}%
\let\auto@bib@innerbib\@empty
\bibitem [{\citenamefont {Bertone}\ \emph {et~al.}(2005)\citenamefont
  {Bertone}, \citenamefont {Hooper},\ and\ \citenamefont
  {Silk}}]{Bertone:2004pz}%
  \BibitemOpen
  \bibfield  {author} {\bibinfo {author} {\bibfnamefont {G.}~\bibnamefont
  {Bertone}}, \bibinfo {author} {\bibfnamefont {D.}~\bibnamefont {Hooper}}, \
  and\ \bibinfo {author} {\bibfnamefont {J.}~\bibnamefont {Silk}},\ }\href@noop
  {} {\bibfield  {journal} {\bibinfo  {journal} {Phys. Rep.}\ }\textbf
  {\bibinfo {volume} {405}},\ \bibinfo {pages} {279} (\bibinfo {year}
  {2005})}\BibitemShut {NoStop}%
\bibitem [{\citenamefont {Gaitskell}(2004)}]{Gaitskell:2004gd}%
  \BibitemOpen
  \bibfield  {author} {\bibinfo {author} {\bibfnamefont {R.~J.}\ \bibnamefont
  {Gaitskell}},\ }\href@noop {} {\bibfield  {journal} {\bibinfo  {journal}
  {Ann. Rev. Nucl. Part. Sci.}\ }\textbf {\bibinfo {volume} {54}},\ \bibinfo
  {pages} {315} (\bibinfo {year} {2004})}\BibitemShut {NoStop}%
\bibitem [{\citenamefont {Jungman}\ \emph {et~al.}(1996)\citenamefont
  {Jungman}, \citenamefont {Kamionkowski},\ and\ \citenamefont
  {Griest}}]{Jungman:1995df}%
  \BibitemOpen
  \bibfield  {author} {\bibinfo {author} {\bibfnamefont {G.}~\bibnamefont
  {Jungman}}, \bibinfo {author} {\bibfnamefont {M.}~\bibnamefont
  {Kamionkowski}}, \ and\ \bibinfo {author} {\bibfnamefont {K.}~\bibnamefont
  {Griest}},\ }\href@noop {} {\bibfield  {journal} {\bibinfo  {journal} {Phys.
  Rep.}\ }\textbf {\bibinfo {volume} {267}},\ \bibinfo {pages} {195} (\bibinfo
  {year} {1996})}\BibitemShut {NoStop}%
\bibitem [{\citenamefont {Bernabei}\ \emph {et~al.}(2008)\citenamefont
  {Bernabei} \emph {et~al.}}]{Bernabei:2008yi}%
  \BibitemOpen
  \bibfield  {author} {\bibinfo {author} {\bibfnamefont {R.}~\bibnamefont
  {Bernabei}} \emph {et~al.} (\bibinfo {collaboration} {DAMA/LIBRA}),\
  }\href@noop {} {\bibfield  {journal} {\bibinfo  {journal} {Eur. Phys. J. C}\
  }\textbf {\bibinfo {volume} {56}},\ \bibinfo {pages} {333} (\bibinfo {year}
  {2008})}\BibitemShut {NoStop}%
\bibitem [{\citenamefont {{Bernabei}}\ \emph {et~al.}(2010)\citenamefont
  {{Bernabei}} \emph {et~al.}}]{Bernabei:2010}%
  \BibitemOpen
  \bibfield  {author} {\bibinfo {author} {\bibfnamefont {R.}~\bibnamefont
  {{Bernabei}}} \emph {et~al.},\ }\href@noop {} {\bibfield  {journal} {\bibinfo
   {journal} {Eur. Phys. J. C}\ }\textbf {\bibinfo {volume} {67}},\ \bibinfo
  {pages} {39} (\bibinfo {year} {2010})}\BibitemShut {NoStop}%
\bibitem [{\citenamefont {Aalseth}\ \emph {et~al.}(2011)\citenamefont {Aalseth}
  \emph {et~al.}}]{Aalseth:2010vx}%
  \BibitemOpen
  \bibfield  {author} {\bibinfo {author} {\bibfnamefont {C.~E.}\ \bibnamefont
  {Aalseth}} \emph {et~al.} (\bibinfo {collaboration} {CoGeNT}),\ }\href@noop
  {} {\bibfield  {journal} {\bibinfo  {journal} {Phys. Rev. Lett.}\ }\textbf
  {\bibinfo {volume} {106}},\ \bibinfo {pages} {131301} (\bibinfo {year}
  {2011})}\BibitemShut {NoStop}%
\bibitem [{\citenamefont {{Aalseth}}\ \emph {et~al.}(2011)\citenamefont
  {{Aalseth}} \emph {et~al.}}]{Aalseth:2011ce}%
  \BibitemOpen
  \bibfield  {author} {\bibinfo {author} {\bibfnamefont {C.~E.}\ \bibnamefont
  {{Aalseth}}} \emph {et~al.},\ }\href@noop {} {\bibfield  {journal} {\bibinfo
  {journal} {Phys. Rev. Lett.}\ }\textbf {\bibinfo {volume} {107}},\ \bibinfo
  {pages} {141301} (\bibinfo {year} {2011})}\BibitemShut {NoStop}%
\bibitem [{\citenamefont {{Angloher}}\ \emph {et~al.}(2011)\citenamefont
  {{Angloher}} \emph {et~al.}}]{Angloher:2011}%
  \BibitemOpen
  \bibfield  {author} {\bibinfo {author} {\bibfnamefont {G.}~\bibnamefont
  {{Angloher}}} \emph {et~al.} (\bibinfo {collaboration} {CRESST-II}),\
  }\href@noop {} {\bibfield  {journal} {\bibinfo  {journal}
  {arXiv:1109.0702v1}\ } (\bibinfo {year} {2011})}\BibitemShut {NoStop}%
\bibitem [{\citenamefont {Hooper}(2012)}]{Hooper:2012fk}%
  \BibitemOpen
  \bibfield  {author} {\bibinfo {author} {\bibfnamefont {D.}~\bibnamefont
  {Hooper}},\ }\href@noop {} {\bibfield  {journal} {\bibinfo  {journal}
  {arXiv:1201.1303v1}\ } (\bibinfo {year} {2012})}\BibitemShut {NoStop}%
\bibitem [{\citenamefont {Fox}\ \emph {et~al.}(2011)\citenamefont {Fox} \emph
  {et~al.}}]{Fox:2011kx}%
  \BibitemOpen
  \bibfield  {author} {\bibinfo {author} {\bibfnamefont {P.~J.}\ \bibnamefont
  {Fox}} \emph {et~al.},\ }\href@noop {} {\bibfield  {journal} {\bibinfo
  {journal} {arXiv:1107.0717v2}\ } (\bibinfo {year} {2011})}\BibitemShut
  {NoStop}%
\bibitem [{\citenamefont {Kopp}\ \emph {et~al.}(2011)\citenamefont {Kopp},
  \citenamefont {Schwetz},\ and\ \citenamefont {Zupan}}]{Kopp:2011uq}%
  \BibitemOpen
  \bibfield  {author} {\bibinfo {author} {\bibfnamefont {J.}~\bibnamefont
  {Kopp}}, \bibinfo {author} {\bibfnamefont {T.}~\bibnamefont {Schwetz}}, \
  and\ \bibinfo {author} {\bibfnamefont {J.}~\bibnamefont {Zupan}},\
  }\href@noop {} {\bibfield  {journal} {\bibinfo  {journal}
  {arXiv:1110.2721v1}\ } (\bibinfo {year} {2011})}\BibitemShut {NoStop}%
\bibitem [{\citenamefont {{Belli}}\ \emph {et~al.}(2011)\citenamefont {{Belli}}
  \emph {et~al.}}]{Belli:2011fa}%
  \BibitemOpen
  \bibfield  {author} {\bibinfo {author} {\bibfnamefont {P.}~\bibnamefont
  {{Belli}}} \emph {et~al.},\ }\href@noop {} {\bibfield  {journal} {\bibinfo
  {journal} {Phys. Rev. D}\ }\textbf {\bibinfo {volume} {84}},\ \bibinfo
  {pages} {055014} (\bibinfo {year} {2011})}\BibitemShut {NoStop}%
\bibitem [{\citenamefont {Drukier}\ \emph {et~al.}(1986)\citenamefont
  {Drukier}, \citenamefont {Freese},\ and\ \citenamefont
  {Spergel}}]{Drukier:1986}%
  \BibitemOpen
  \bibfield  {author} {\bibinfo {author} {\bibfnamefont {A.~K.}\ \bibnamefont
  {Drukier}}, \bibinfo {author} {\bibfnamefont {K.}~\bibnamefont {Freese}}, \
  and\ \bibinfo {author} {\bibfnamefont {D.~N.}\ \bibnamefont {Spergel}},\
  }\href@noop {} {\bibfield  {journal} {\bibinfo  {journal} {Phys. Rev. D}\
  }\textbf {\bibinfo {volume} {33}},\ \bibinfo {pages} {3495} (\bibinfo {year}
  {1986})}\BibitemShut {NoStop}%
\bibitem [{\citenamefont {Akerib}\ \emph {et~al.}(2005)\citenamefont {Akerib}
  \emph {et~al.}}]{Akerib:2005zy}%
  \BibitemOpen
  \bibfield  {author} {\bibinfo {author} {\bibfnamefont {D.~S.}\ \bibnamefont
  {Akerib}} \emph {et~al.} (\bibinfo {collaboration} {CDMS}),\ }\href@noop {}
  {\bibfield  {journal} {\bibinfo  {journal} {Phys. Rev. D}\ }\textbf {\bibinfo
  {volume} {72}},\ \bibinfo {pages} {052009} (\bibinfo {year}
  {2005})}\BibitemShut {NoStop}%
\bibitem [{\citenamefont {Ahmed}\ \emph {et~al.}(2009)\citenamefont {Ahmed}
  \emph {et~al.}}]{Ahmed:2008eu}%
  \BibitemOpen
  \bibfield  {author} {\bibinfo {author} {\bibfnamefont {Z.}~\bibnamefont
  {Ahmed}} \emph {et~al.} (\bibinfo {collaboration} {CDMS}),\ }\href@noop {}
  {\bibfield  {journal} {\bibinfo  {journal} {Phys. Rev. Lett.}\ }\textbf
  {\bibinfo {volume} {102}},\ \bibinfo {pages} {011301} (\bibinfo {year}
  {2009})}\BibitemShut {NoStop}%
\bibitem [{\citenamefont {Ahmed}\ \emph {et~al.}(2010)\citenamefont {Ahmed}
  \emph {et~al.}}]{CDMSScience:2010}%
  \BibitemOpen
  \bibfield  {author} {\bibinfo {author} {\bibfnamefont {Z.}~\bibnamefont
  {Ahmed}} \emph {et~al.} (\bibinfo {collaboration} {CDMS}),\ }\href@noop {}
  {\bibfield  {journal} {\bibinfo  {journal} {Science}\ }\textbf {\bibinfo
  {volume} {327}},\ \bibinfo {pages} {1619} (\bibinfo {year}
  {2010})}\BibitemShut {NoStop}%
\bibitem [{\citenamefont {Ahmed}\ \emph {et~al.}(2011)\citenamefont {Ahmed}
  \emph {et~al.}}]{Ahmed:2011le}%
  \BibitemOpen
  \bibfield  {author} {\bibinfo {author} {\bibfnamefont {Z.}~\bibnamefont
  {Ahmed}} \emph {et~al.} (\bibinfo {collaboration} {CDMS}),\ }\href@noop {}
  {\bibfield  {journal} {\bibinfo  {journal} {Phys. Rev. Lett.}\ }\textbf
  {\bibinfo {volume} {106}},\ \bibinfo {pages} {131302} (\bibinfo {year}
  {2011})},\ \Eprint {http://arxiv.org/abs/arXiv:1011.2482v3}
  {arXiv:1011.2482v3} \BibitemShut {NoStop}%
\bibitem [{\citenamefont {Collar}()}]{Collar:TAUP}%
  \BibitemOpen
  \bibfield  {author} {\bibinfo {author} {\bibfnamefont {J.~I.}\ \bibnamefont
  {Collar}},\ }\href@noop {} {}\bibinfo {note} {Talk at TAUP 2011 Workshop,
  Munich, Germany, Sep. 5-9, 2011}\BibitemShut {NoStop}%
\bibitem [{\citenamefont {Kelso}\ \emph {et~al.}(2011)\citenamefont {Kelso},
  \citenamefont {Hooper},\ and\ \citenamefont {Buckley}}]{Kelso:2011vn}%
  \BibitemOpen
  \bibfield  {author} {\bibinfo {author} {\bibfnamefont {C.}~\bibnamefont
  {Kelso}}, \bibinfo {author} {\bibfnamefont {D.}~\bibnamefont {Hooper}}, \
  and\ \bibinfo {author} {\bibfnamefont {M.~R.}\ \bibnamefont {Buckley}},\
  }\href@noop {} {\bibfield  {journal} {\bibinfo  {journal}
  {arXiv:1110.5338v1}\ } (\bibinfo {year} {2011})}\BibitemShut {NoStop}%
\bibitem [{\citenamefont {{Schwetz}}\ and\ \citenamefont
  {{Zupan}}(2011)}]{Schwetz:2011}%
  \BibitemOpen
  \bibfield  {author} {\bibinfo {author} {\bibfnamefont {T.}~\bibnamefont
  {{Schwetz}}}\ and\ \bibinfo {author} {\bibfnamefont {J.}~\bibnamefont
  {{Zupan}}},\ }\href {\doibase 10.1088/1475-7516/2011/08/008} {\bibfield
  {journal} {\bibinfo  {journal} {JCAP}\ }\textbf {\bibinfo {volume} {8}},\
  \bibinfo {pages} {8} (\bibinfo {year} {2011})},\ \Eprint
  {http://arxiv.org/abs/1106.6241} {arXiv:1106.6241 [hep-ph]} \BibitemShut
  {NoStop}%
\bibitem [{\citenamefont {{Farina}}\ \emph {et~al.}(2011)\citenamefont
  {{Farina}}, \citenamefont {{Pappadopulo}}, \citenamefont {{Strumia}},\ and\
  \citenamefont {{Volansky}}}]{Farina:2011}%
  \BibitemOpen
  \bibfield  {author} {\bibinfo {author} {\bibfnamefont {M.}~\bibnamefont
  {{Farina}}}, \bibinfo {author} {\bibfnamefont {D.}~\bibnamefont
  {{Pappadopulo}}}, \bibinfo {author} {\bibfnamefont {A.}~\bibnamefont
  {{Strumia}}}, \ and\ \bibinfo {author} {\bibfnamefont {T.}~\bibnamefont
  {{Volansky}}},\ }\href {\doibase 10.1088/1475-7516/2011/11/010} {\bibfield
  {journal} {\bibinfo  {journal} {JCAP}\ }\textbf {\bibinfo {volume} {11}},\
  \bibinfo {pages} {10} (\bibinfo {year} {2011})},\ \Eprint
  {http://arxiv.org/abs/1107.0715} {arXiv:1107.0715 [hep-ph]} \BibitemShut
  {NoStop}%
\bibitem [{\citenamefont {Lindhard}\ \emph {et~al.}(1963)\citenamefont
  {Lindhard} \emph {et~al.}}]{Lindhard:1963}%
  \BibitemOpen
  \bibfield  {author} {\bibinfo {author} {\bibfnamefont {J.}~\bibnamefont
  {Lindhard}} \emph {et~al.},\ }\href@noop {} {\bibfield  {journal} {\bibinfo
  {journal} {K. Dan. Vidensk. Selsk., Mat.-Fys. Medd.}\ }\textbf {\bibinfo
  {volume} {33}},\ \bibinfo {pages} {10} (\bibinfo {year} {1963})}\BibitemShut
  {NoStop}%
\bibitem [{\citenamefont {Lewin}\ and\ \citenamefont
  {Smith}(1996)}]{Lewin:1995rx}%
  \BibitemOpen
  \bibfield  {author} {\bibinfo {author} {\bibfnamefont {J.~D.}\ \bibnamefont
  {Lewin}}\ and\ \bibinfo {author} {\bibfnamefont {P.~F.}\ \bibnamefont
  {Smith}},\ }\href@noop {} {\bibfield  {journal} {\bibinfo  {journal}
  {Astropart. Phys.}\ }\textbf {\bibinfo {volume} {6}},\ \bibinfo {pages} {87}
  (\bibinfo {year} {1996})}\BibitemShut {NoStop}%
\bibitem [{\citenamefont {Collar}(2011)}]{Collar:priv}%
  \BibitemOpen
  \bibfield  {author} {\bibinfo {author} {\bibfnamefont {J.}~\bibnamefont
  {Collar}},\ }\href@noop {} {\bibfield  {journal} {\bibinfo  {journal}
  {private communication}\ } (\bibinfo {year} {2011})}\BibitemShut {NoStop}%
\bibitem [{\citenamefont {Angle}\ \emph {et~al.}(2011)\citenamefont {Angle}
  \emph {et~al.}}]{Angle:2011}%
  \BibitemOpen
  \bibfield  {author} {\bibinfo {author} {\bibfnamefont {J.}~\bibnamefont
  {Angle}} \emph {et~al.} (\bibinfo {collaboration} {XENON10}),\ }\href@noop {}
  {\bibfield  {journal} {\bibinfo  {journal} {Phys. Rev. Lett.}\ }\textbf
  {\bibinfo {volume} {107}},\ \bibinfo {pages} {051301} (\bibinfo {year}
  {2011})}\BibitemShut {NoStop}%
\bibitem [{\citenamefont {Aprile}\ \emph {et~al.}(2011)\citenamefont {Aprile}
  \emph {et~al.}}]{Aprile:2011ba}%
  \BibitemOpen
  \bibfield  {author} {\bibinfo {author} {\bibfnamefont {E.}~\bibnamefont
  {Aprile}} \emph {et~al.} (\bibinfo {collaboration} {XENON100}),\ }\href@noop
  {} {\bibfield  {journal} {\bibinfo  {journal} {Phys. Rev. Lett.}\ }\textbf
  {\bibinfo {volume} {107}},\ \bibinfo {pages} {131302} (\bibinfo {year}
  {2011})}\BibitemShut {NoStop}%
\bibitem [{\citenamefont {Neganov}\ and\ \citenamefont
  {Trofimov}(1985)}]{Neganov:1985}%
  \BibitemOpen
  \bibfield  {author} {\bibinfo {author} {\bibfnamefont {B.}~\bibnamefont
  {Neganov}}\ and\ \bibinfo {author} {\bibfnamefont {V.}~\bibnamefont
  {Trofimov}},\ }\href@noop {} {\bibfield  {journal} {\bibinfo  {journal}
  {Otkryt. Izobret.}\ }\textbf {\bibinfo {volume} {146}},\ \bibinfo {pages}
  {215} (\bibinfo {year} {1985})}\BibitemShut {NoStop}%
\bibitem [{\citenamefont {{Luke}}(1988)}]{Luke:1988}%
  \BibitemOpen
  \bibfield  {author} {\bibinfo {author} {\bibfnamefont {P.~N.}\ \bibnamefont
  {{Luke}}},\ }\href@noop {} {\bibfield  {journal} {\bibinfo  {journal} {J.
  Appl. Phys.}\ }\textbf {\bibinfo {volume} {64}},\ \bibinfo {pages} {6858}
  (\bibinfo {year} {1988})}\BibitemShut {NoStop}%
\bibitem [{\citenamefont {Barbeau}(2009)}]{Barbeau:thesis}%
  \BibitemOpen
  \bibfield  {author} {\bibinfo {author} {\bibfnamefont {P.}~\bibnamefont
  {Barbeau}},\ }\href@noop {} {\bibfield  {journal} {\bibinfo  {journal} {Ph.D.
  thesis, U. Chicago}\ } (\bibinfo {year} {2009})}\BibitemShut {NoStop}%
\bibitem [{\citenamefont {Feldman}\ and\ \citenamefont
  {Cousins}(1998)}]{Feldman:1998}%
  \BibitemOpen
  \bibfield  {author} {\bibinfo {author} {\bibfnamefont {G.~J.}\ \bibnamefont
  {Feldman}}\ and\ \bibinfo {author} {\bibfnamefont {R.~D.}\ \bibnamefont
  {Cousins}},\ }\href {\doibase 10.1103/PhysRevD.57.3873} {\bibfield  {journal}
  {\bibinfo  {journal} {Phys. Rev. D}\ }\textbf {\bibinfo {volume} {57}},\
  \bibinfo {pages} {3873} (\bibinfo {year} {1998})}\BibitemShut {NoStop}%
\bibitem [{\citenamefont {{Navick}}\ \emph {et~al.}(2000)\citenamefont
  {{Navick}} \emph {et~al.}}]{Navick:2000}%
  \BibitemOpen
  \bibfield  {author} {\bibinfo {author} {\bibfnamefont {X.-F.}\ \bibnamefont
  {{Navick}}} \emph {et~al.},\ }\href {\doibase 10.1016/S0168-9002(99)01394-7}
  {\bibfield  {journal} {\bibinfo  {journal} {NIM A}\ }\textbf {\bibinfo
  {volume} {444}},\ \bibinfo {pages} {361} (\bibinfo {year}
  {2000})}\BibitemShut {NoStop}%
\bibitem [{\citenamefont {{Shutt}}\ \emph {et~al.}(1992)\citenamefont {{Shutt}}
  \emph {et~al.}}]{Shutt:1992}%
  \BibitemOpen
  \bibfield  {author} {\bibinfo {author} {\bibfnamefont {T.}~\bibnamefont
  {{Shutt}}} \emph {et~al.},\ }\href {\doibase 10.1103/PhysRevLett.69.3531}
  {\bibfield  {journal} {\bibinfo  {journal} {Physical Review Letters}\
  }\textbf {\bibinfo {volume} {69}},\ \bibinfo {pages} {3531} (\bibinfo {year}
  {1992})}\BibitemShut {NoStop}%
\bibitem [{\citenamefont {Akerib}\ \emph {et~al.}(2010)\citenamefont {Akerib}
  \emph {et~al.}}]{Akerib:2010rr}%
  \BibitemOpen
  \bibfield  {author} {\bibinfo {author} {\bibfnamefont {D.~S.}\ \bibnamefont
  {Akerib}} \emph {et~al.} (\bibinfo {collaboration} {CDMS}),\ }\href@noop {}
  {\bibfield  {journal} {\bibinfo  {journal} {Phys. Rev. D}\ }\textbf {\bibinfo
  {volume} {82}},\ \bibinfo {pages} {122004} (\bibinfo {year}
  {2010})}\BibitemShut {NoStop}%
\bibitem [{\citenamefont {Hooper}\ \emph {et~al.}(2010)\citenamefont {Hooper}
  \emph {et~al.}}]{Hooper:2010ly}%
  \BibitemOpen
  \bibfield  {author} {\bibinfo {author} {\bibfnamefont {D.}~\bibnamefont
  {Hooper}} \emph {et~al.},\ }\href@noop {} {\bibfield  {journal} {\bibinfo
  {journal} {Phys. Rev. D}\ }\textbf {\bibinfo {volume} {82}},\ \bibinfo
  {pages} {123509} (\bibinfo {year} {2010})}\BibitemShut {NoStop}%
\bibitem [{\citenamefont {{Fox}}\ \emph {et~al.}(2011)\citenamefont {{Fox}},
  \citenamefont {{Liu}},\ and\ \citenamefont
  {{Weiner}}}]{Fox:2011integrateout}%
  \BibitemOpen
  \bibfield  {author} {\bibinfo {author} {\bibfnamefont {P.~J.}\ \bibnamefont
  {{Fox}}}, \bibinfo {author} {\bibfnamefont {J.}~\bibnamefont {{Liu}}}, \ and\
  \bibinfo {author} {\bibfnamefont {N.}~\bibnamefont {{Weiner}}},\ }\href
  {\doibase 10.1103/PhysRevD.83.103514} {\bibfield  {journal} {\bibinfo
  {journal} {Phys. Rev. D}\ }\textbf {\bibinfo {volume} {83}},\ \bibinfo {eid}
  {103514} (\bibinfo {year} {2011})},\ \Eprint
  {http://arxiv.org/abs/arXiv:1011.1915} {arXiv:1011.1915} \BibitemShut
  {NoStop}%
\end{thebibliography}%
\bibliographystyle{apsrev4-1}

\clearpage
\onecolumngrid
\appendix
\section{Supplementary Material}

The following pages provide relevant additional information.

\vspace{10mm}
 
\subsection{Energy Scales}

The total energy deposited by a particle interaction is the ``recoil energy,'' $E_R$.  The majority of the recoil energy is deposited directly as athermal phonons, while a fraction of the energy goes into producing electron-hole pairs above the band gap.  After the charge carriers are drifted across the detector and relax to the Fermi level, the remaining recoil energy is deposited in the phonon system.

On average one electron-hole pair is produced for every $\epsilon = 3.0$\,eV of recoil energy for an electron recoil in Ge~\cite{Navick:2000,Shutt:1992}.  The ``ionization energy'' $E_Q$ is defined for convenience as the recoil energy inferred from the detected number of charge pairs, $N_Q$, assuming that the event is an electron recoil:
\begin{equation}
E_Q \equiv N_Q \times \epsilon .
\end{equation}

The ionization energy scale is calibrated \emph{in situ} using electron-recoil lines of known energy, and is reported in units of ``keV$_{\mathrm{ee}}$,'' which gives the recoil energy in keV for an electron recoil producing the same ionization signal. The ionization yield $Y \equiv E_Q / E_R$, so on average $Y = 1$ for electron recoils.  Nuclear recoils produce fewer charge pairs, and hence less ionization energy, $E_Q$, than electron recoils of the same recoil energy. The ionization yield $Y$ for nuclear-recoil events depends on recoil energy, with $Y \approx 0.2$\textendash0.25 in Ge for $E_R \approx 5$\textendash10\,keV~\cite{Lindhard:1963,Lewin:1995rx}. 

An additional phonon population is produced by drifting the charge carriers across the detector, with the total energy deposited equal to the work done by the electric field.  These ``Neganov-Luke'' phonons~\cite{Neganov:1985,*Luke:1988} contribute to the total observed phonon signal, $E_P$, yielding
\begin{equation}
\label{eq:tot_phonon}
E_P = E_R + e V_b N_Q 
               = E_R + \frac{e V_b}{\epsilon}E_Q  ,
\end{equation}
where $V_b = 3.0$\,V is the bias voltage across the detector. Since $E_Q = E_R$ for electron recoils with full charge collection, $E_P = \left( 1 + \frac{e V_b}{\epsilon} \right) E_R$ for these events.

In this, as in previous low-energy analyses~\cite{Ahmed:2011le,Akerib:2010rr}, the recoil energy for each event is determined from the phonon signal alone, after accounting for the Neganov-Luke phonons using Eq.~\ref{eq:tot_phonon}.  This recoil energy estimate avoids incorporating the poorer signal-to-noise of the charge measurement at low energy, but requires an assumption of the recoil type to correctly estimate the recoil energy.

For electron recoils, the phonon-based recoil energy scale, $E_{ER}(E_P)$, is directly calibrated \emph{in situ} using the 1.3\,\kevee and 10.4\,\kevee Ge activation lines visible after neutron calibrations, ensuring that this energy scale cannot be overestimated at the 90\% C.L.  Due to the calibration procedure, this scale is given by definition as
\begin{equation}
E_{ER}(E_P) = \frac{E_P}{2}  ,
\end{equation}
where we have used that $eV_b/\epsilon = 1$ when biased at the calibrated voltage $V_b = 3.0$\,V.  

The energy scale for nuclear recoils, $E_{NR}(E_P)$, reported in units of ``keV$_{\mathrm{nr}}$,'' uses the same calibration of the phonon energy scale as for electron recoils, but with a smaller correction for the Neganov-Luke phonons produced by a nuclear recoil.  The mean ionization energy for nuclear recoils $E_{Q,NR}(E_P)$ is determined over the relevant energy range from the distribution of nuclear recoils from \emph{in situ} \cf calibration data. The nuclear recoil energy scale is then
\begin{equation}
E_{NR}(E_P) = E_P - E_{Q,NR}(E_P)   .
\end{equation}

As stated above, the electron-recoil energy scale $E_{ER}(E_P)$ is directly calibrated with lines of known energy.  The nuclear recoil energy scale cannot be directly calibrated in a similar fashion due to the lack of spectral features in the energy range of interest.  Errors in the nuclear-recoil energy scale could be introduced by errors in the Neganov-Luke phonon correction relative to electron recoils, or dependence of the athermal phonon collection on recoil type.  Since the Neganov-Luke phonon contribution accounts for only 20\% of the total phonon signal and the ionization produced by nuclear recoils is well-measured using the \cf calibration data, the correction for the Neganov-Luke phonons does not contribute significant uncertainty to the recoil energy scale~\cite{Ahmed:2011le}. The absolute nuclear recoil energy scale is constrained by comparison of the measured ionization yield in CDMS to previous measurements of the ionization yield for nuclear recoils of known recoil energy (see e.g.~\cite{Hooper:2010ly} and references therein).  The measured yields are inconsistent with an overestimate of the nuclear recoil energy scale in CDMS, assuming the same ionization collection efficiency as previous measurements~\cite{Ahmed:2011le}.

CoGeNT must similarly assume either a nuclear-recoil energy scale or an electron-recoil energy scale.  In this analysis, we follow the prescription~\cite{Aalseth:2010vx} of the CoGeNT collaboration to relate the ionization energy (calibrated using electron-recoils) to the nuclear recoil energy
\begin{equation}
E_{\mathrm{Q}} = (0.19935)E_{\mathrm{NR}}^{(1.1204)}   ,
\end{equation}
good over the energy range 0.2~keV$_{\mathrm{nr}}< E_{\mathrm{NR}} \lesssim$10~keV$_{\mathrm{nr}}$.

\vspace{20mm}

\subsection{The CDMS II Exposure}
\begin{figure}[htbp]
\begin{center}
\includegraphics[width=7in]{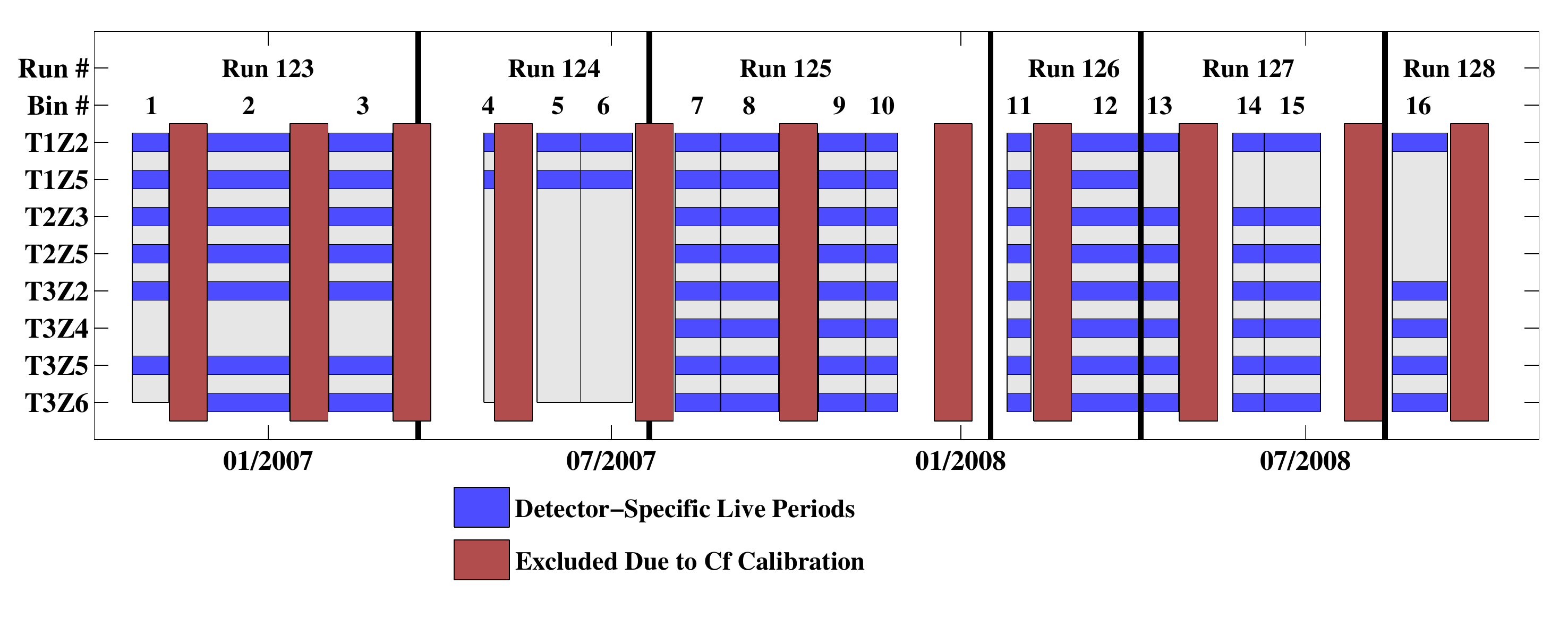}
\caption{\small 
The CDMS II exposure, displaying the detectors and time bins used in this analysis.  For each time bin, a detector is colored blue if this detector's data was used in this analysis.  Divisions between ``runs'' represent at least partial warm-ups of the dilution refrigerator used to cool the detectors.  In order to avoid the effects of Ge activation, 20-day periods were omitted (red) following each \cf calibration time.}
\label{fig:livetimeplot}
\end{center}
\end{figure}

As noted in the main text, the CDMS WIMP-search data are not continuous over the nearly two years of exposure considered here, but include gaps due to neutron calibrations, warming of the detectors, and data periods removed due to data-quality diagnostics.  Since data-quality diagnostics remove individual detectors, different detectors serve as WIMP-search detectors for different times. Figure~\ref{fig:livetimeplot} shows the history of exposure for the period considered on a detector-by-detector basis.  The detectors were arranged in five ``towers,'' and are identified by their tower number (T1\textendash T5) and by their (top-to-bottom) ordering within the tower (Z1\textendash Z6).  
 
\clearpage

\subsection{Event Selection}

\begin{figure}[htbp]
\begin{center}
\includegraphics[width=5.5in]{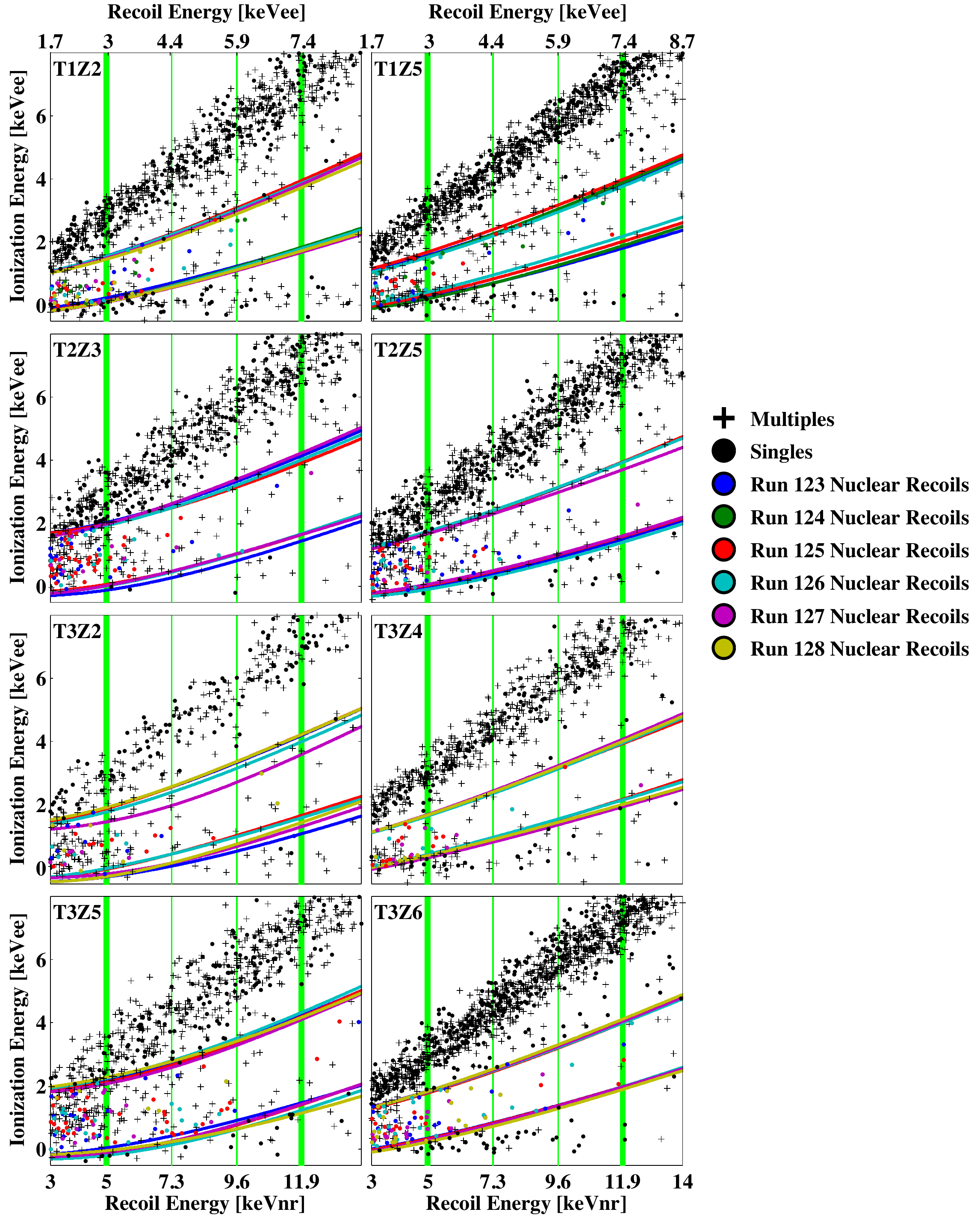}
\caption{\small 
Event selection for the eight detectors employed by this analysis, shown in an ionization-energy vs. phonon-energy plane.  The energy range of this analysis is indicated by thick green lines, along with borders between smaller energy bins.  In this and following plots, the 5\textendash11.9\,keV$_{\mathrm{nr}}$ range has been subdivided into three equal parts: [5\textendash7.3\,keV$_{\mathrm{nr}}$], [7.3\textendash9.6\,keV$_{\mathrm{nr}}$], and [9.6\textendash11.9\,keV$_{\mathrm{nr}}$].  All events shown have passed data-quality cuts, as well as the veto cut and the $Q$-inner cut.  Crosses represent events registering in multiple detectors (``multiples''), filled markers represent events registering in only one detector (``singles''), and colored markers represent singles lying within the $\pm 2\sigma$ nuclear-recoil bands, defined through \cf calibration independently for each detector and run.  The edges of these run-by-run nuclear-recoil band definitions are also indicated.}
\label{fig:yield}
\end{center}
\end{figure}

\clearpage

\subsection{Limits on Modulation of Efficiencies}

\vspace{20mm}

\begin{figure}[htbp]
\begin{center}
\includegraphics[width=3.3in]{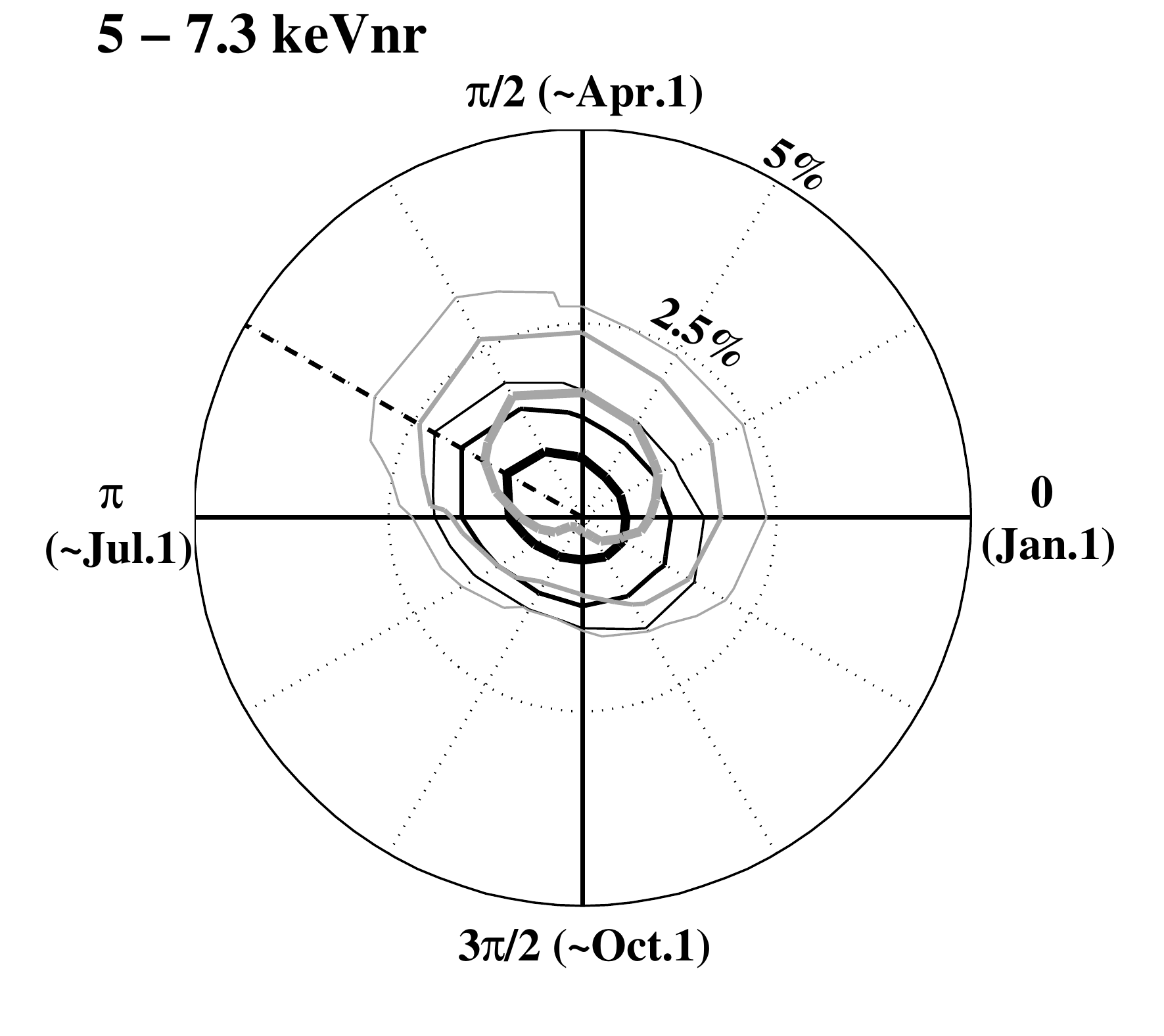}
\includegraphics[width=3.3in]{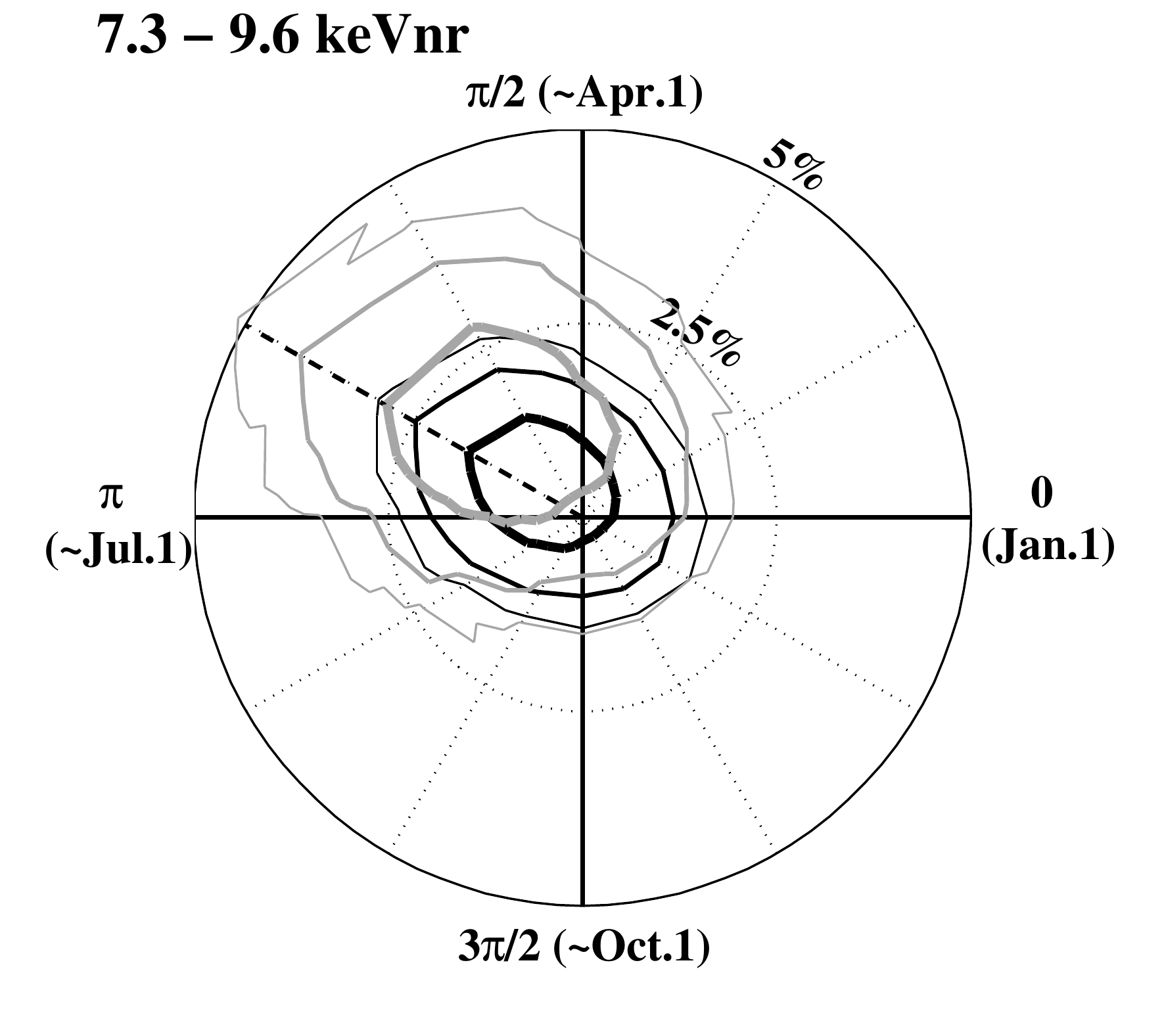}
\includegraphics[width=3.3in]{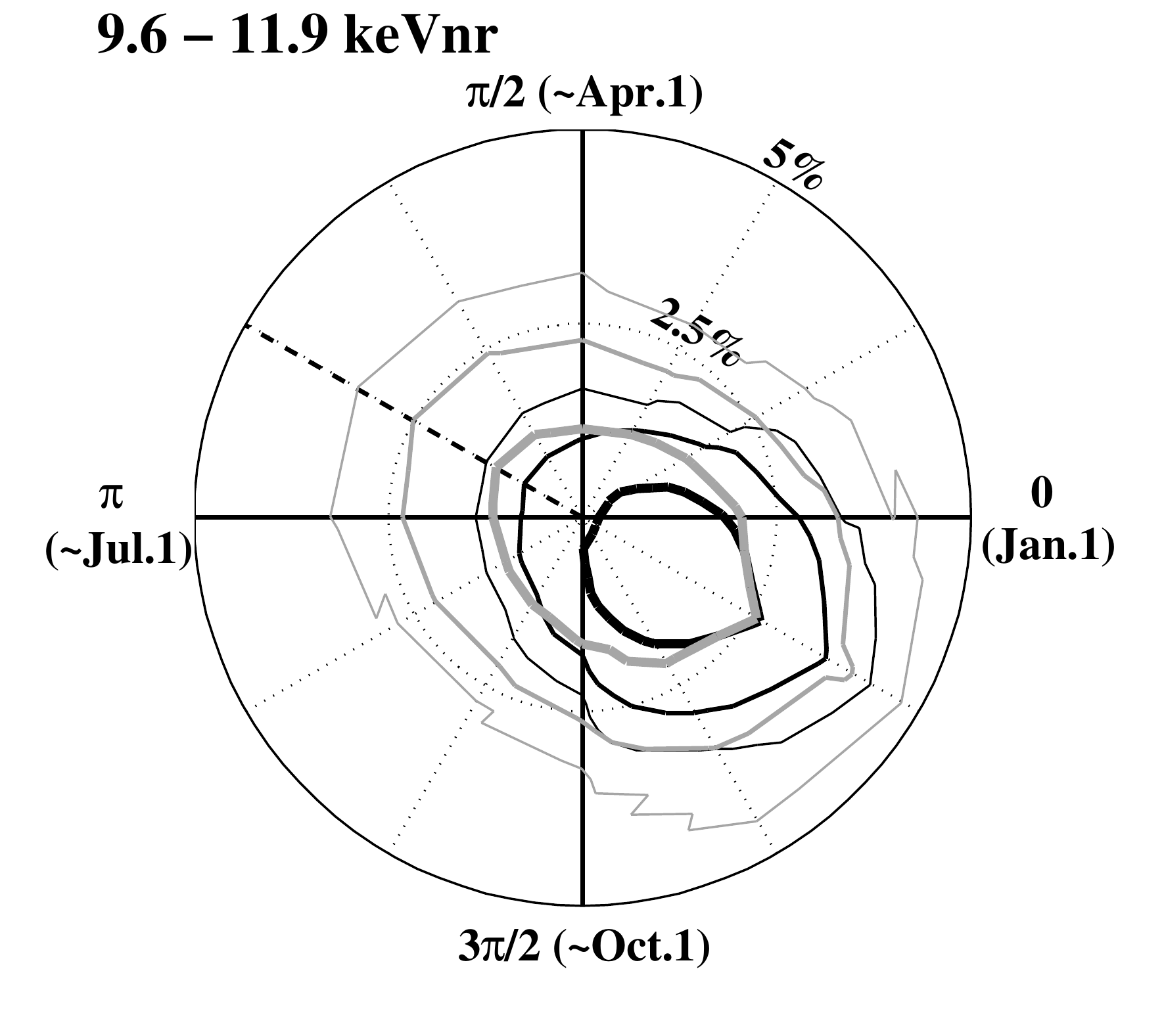}
\includegraphics[width=3.3in]{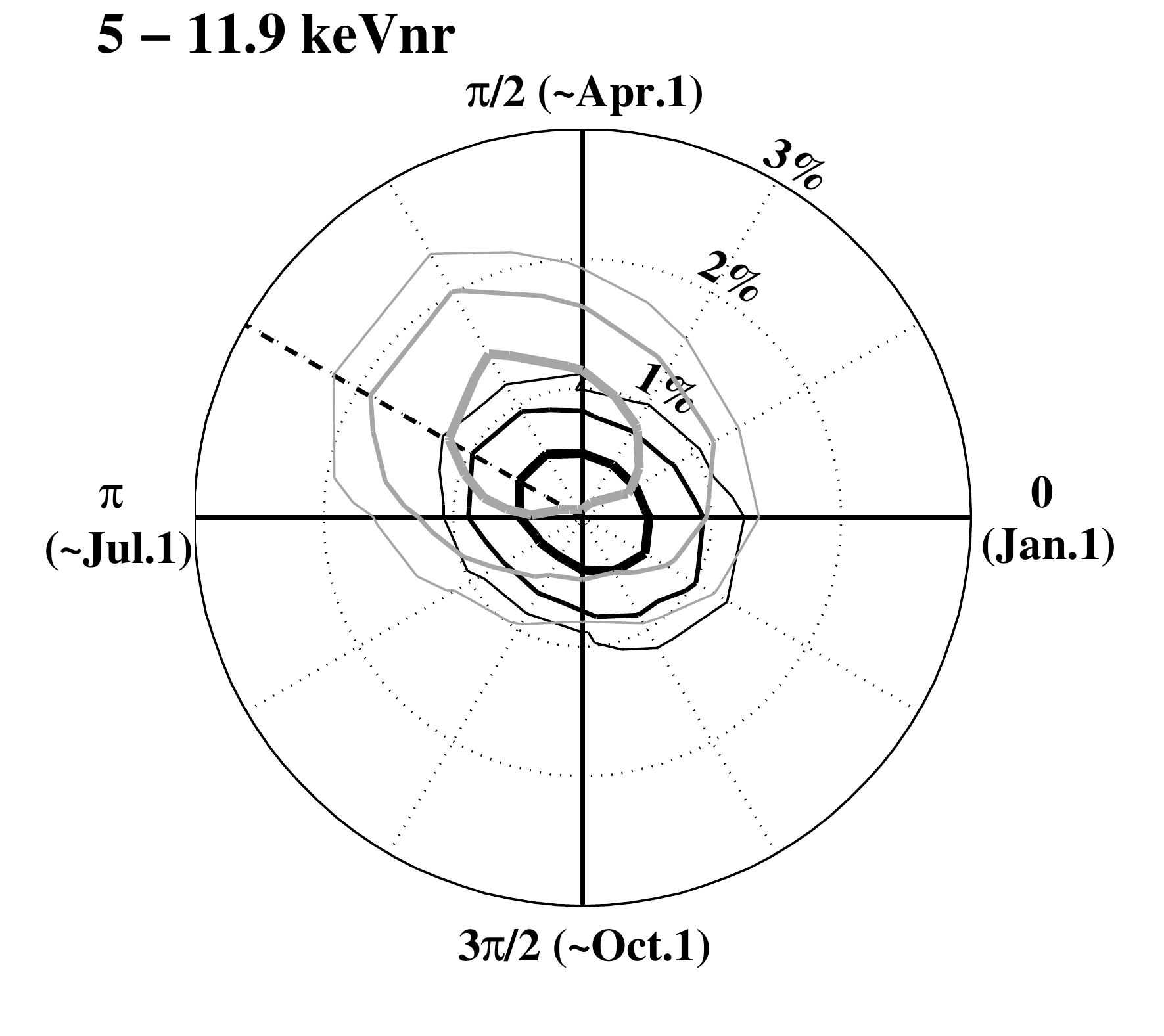}
\caption{\small 
Confidence limits on the relative amplitude and phase of annual modulation in the nuclear-recoil band cut efficiency (black) and the $Q$-inner cut efficiency (grey). Three different energy bins are shown, along with the total energy range (lower right).  Contours are 68, 95, and 99$\%$.}
\label{fig:NR_rmodVphase}
\end{center}
\end{figure}

\clearpage

\subsection{Rates vs. time}

\vspace{10mm}

\begin{figure}[htbp]
\begin{center}
\includegraphics[width=2.4in]{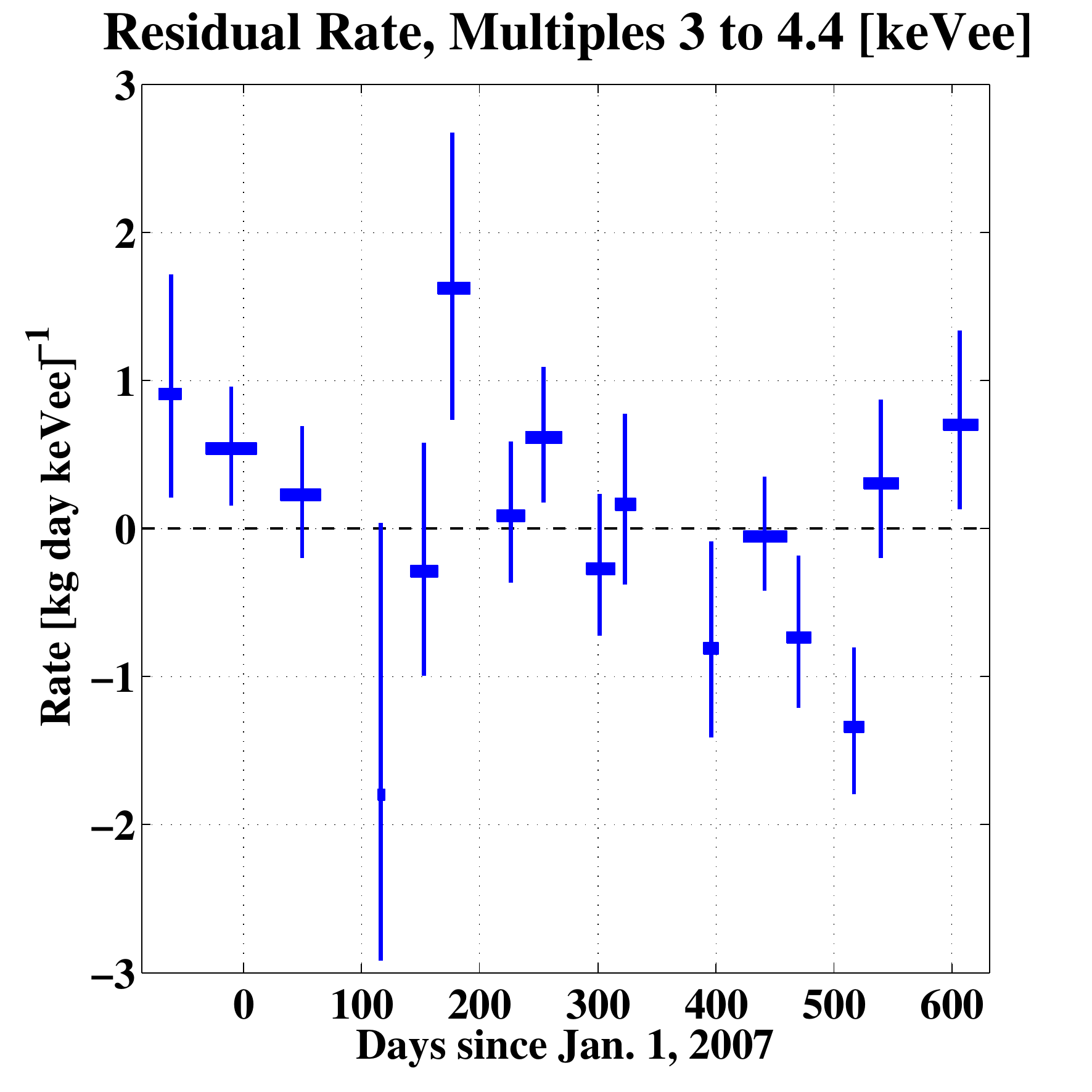}
\includegraphics[width=2.4in]{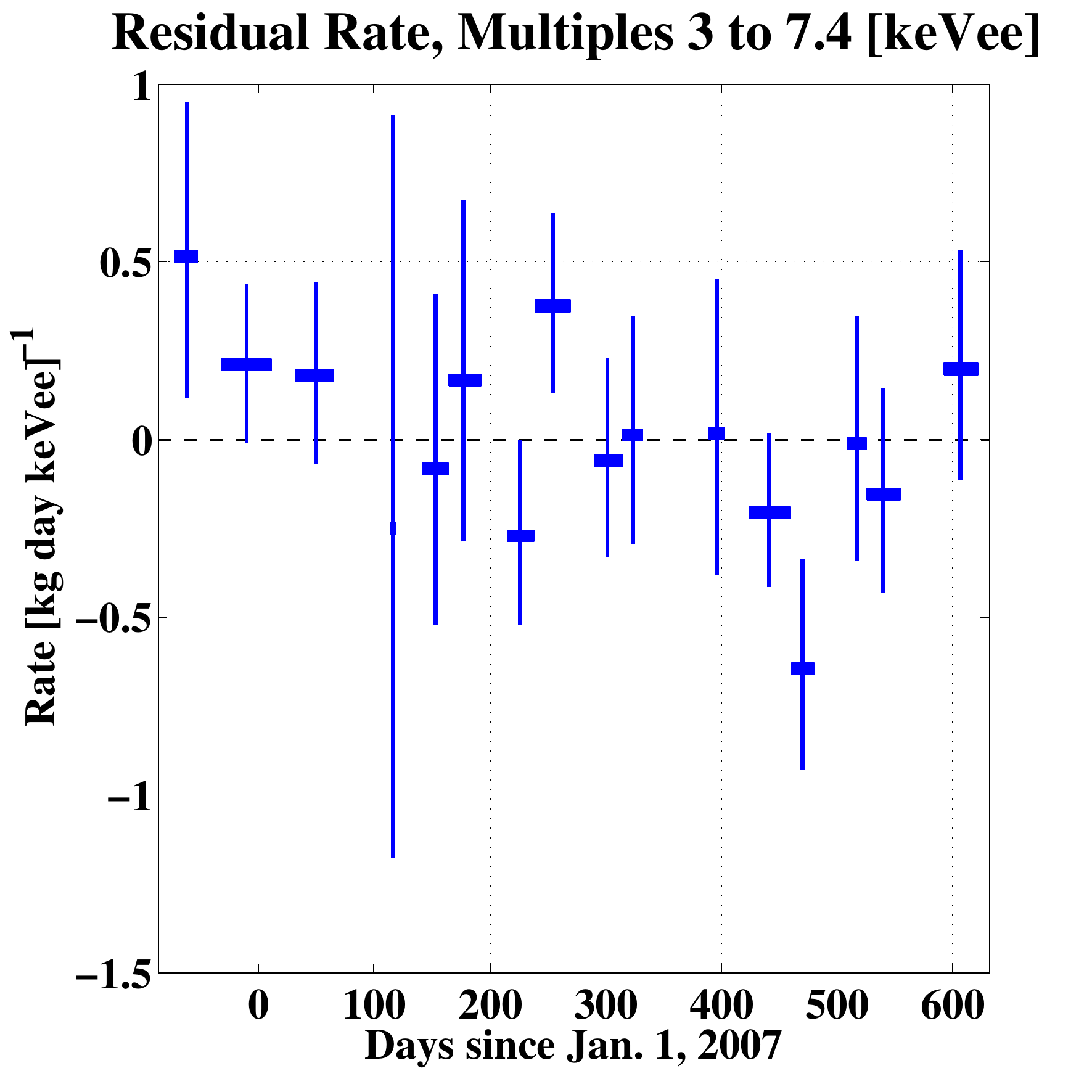}
\includegraphics[width=2.4in]{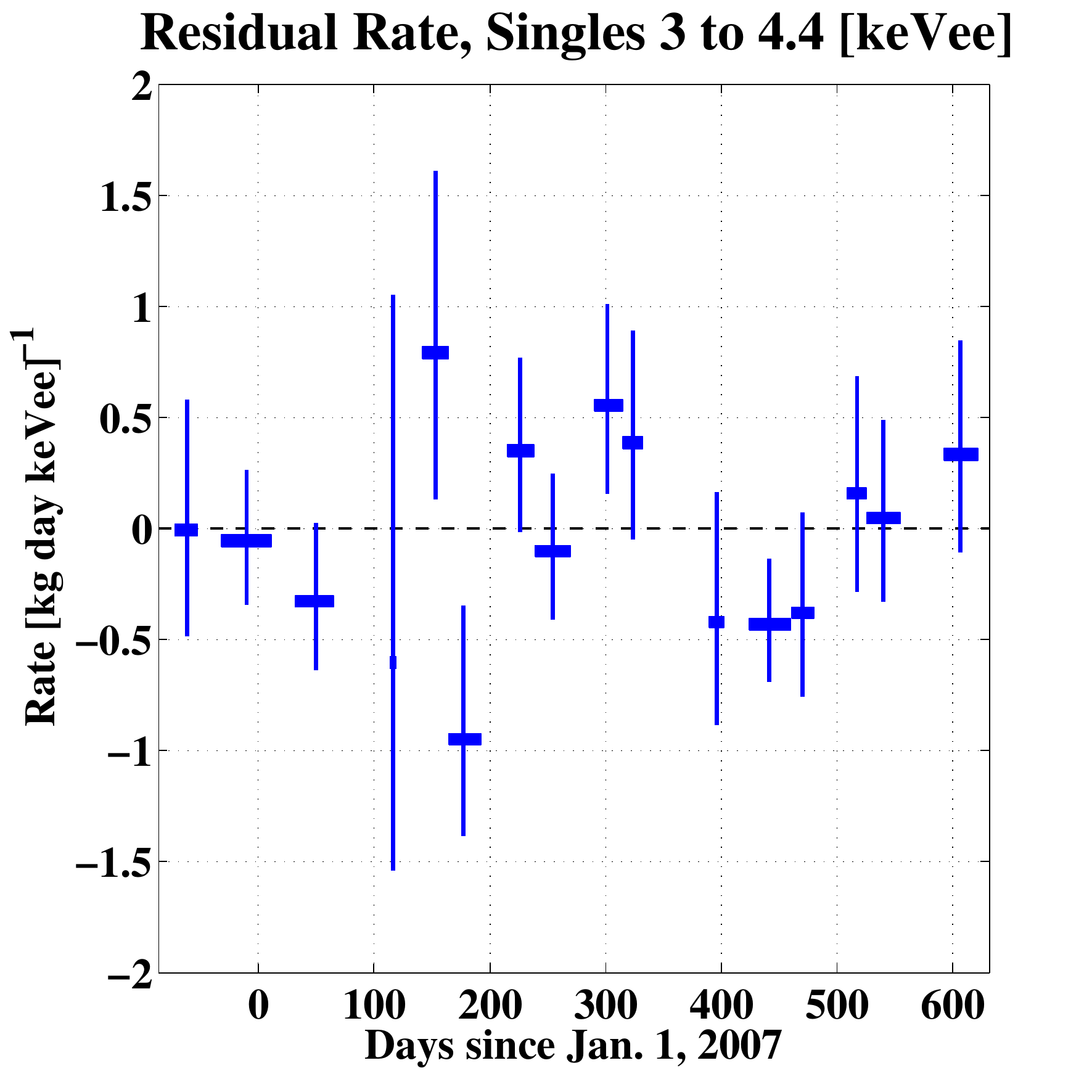}
\includegraphics[width=2.4in]{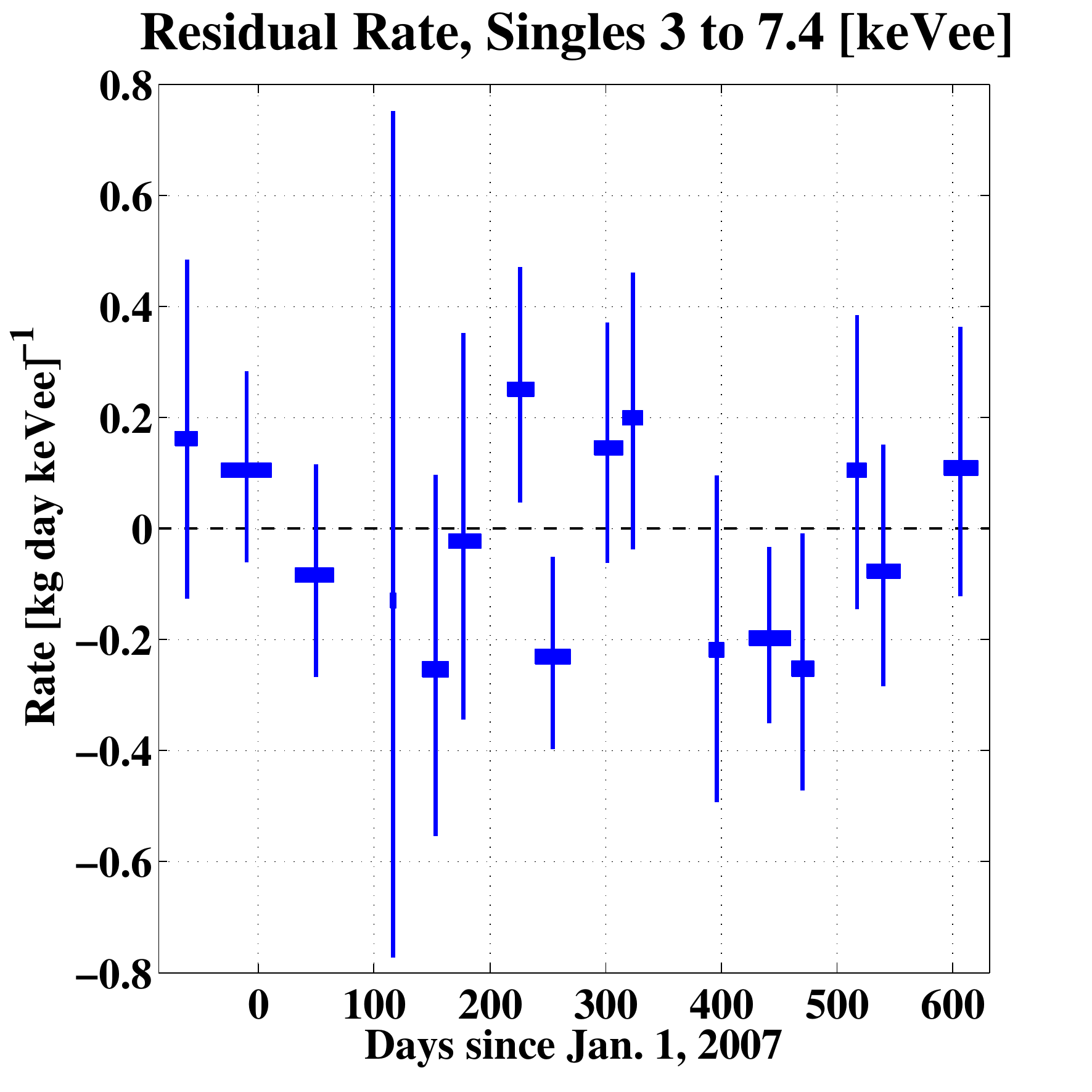}
\includegraphics[width=2.4in]{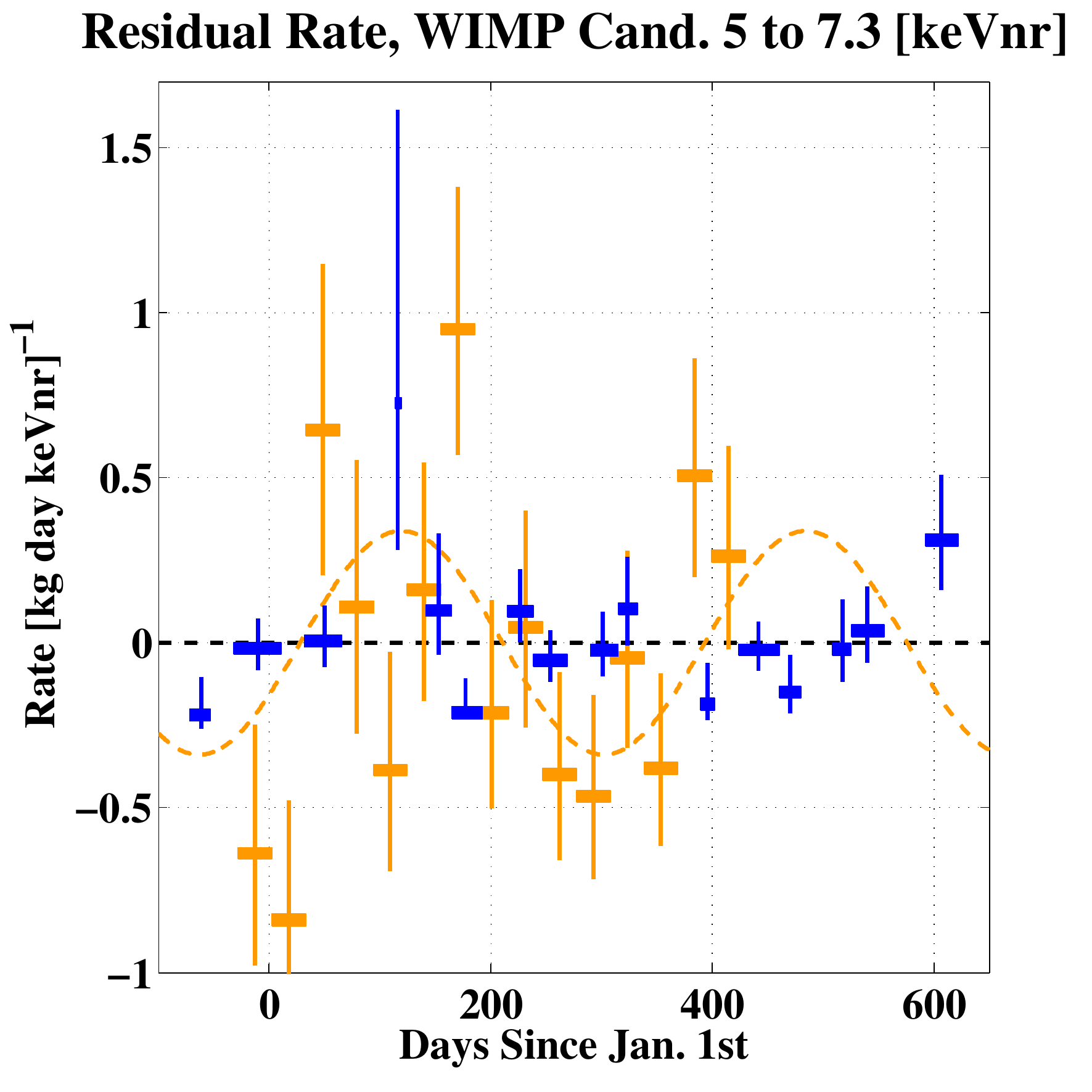}
\includegraphics[width=2.4in]{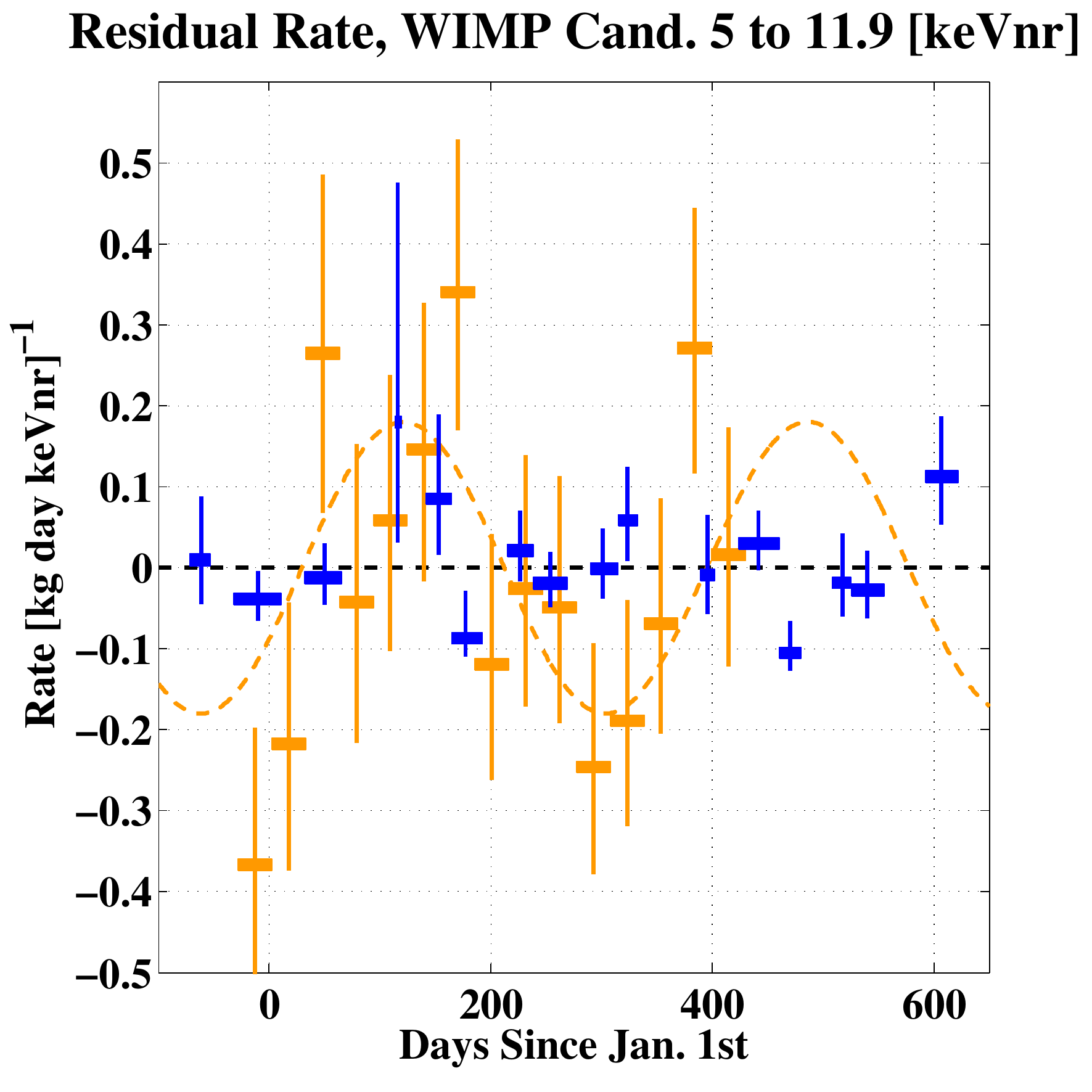}
\caption{\small 
Residual event rate as a function of time, for three event populations:  multiples (top), singles (middle), and singles in the nuclear-recoil band (bottom), as defined in the text.  Two energy ranges are shown, in the left and right columns.  Because the multiples and singles populations are dominated by electron recoils, an electron-recoil energy scale has been used for these rates.}
\label{fig:ratevstime}
\end{center}
\end{figure}

\clearpage
\subsection{Modulation Spectra:  Nuclear-Recoil Singles}
\vspace{20mm}

\begin{figure}[htbp]
\begin{center}
\includegraphics[width=3.5in]{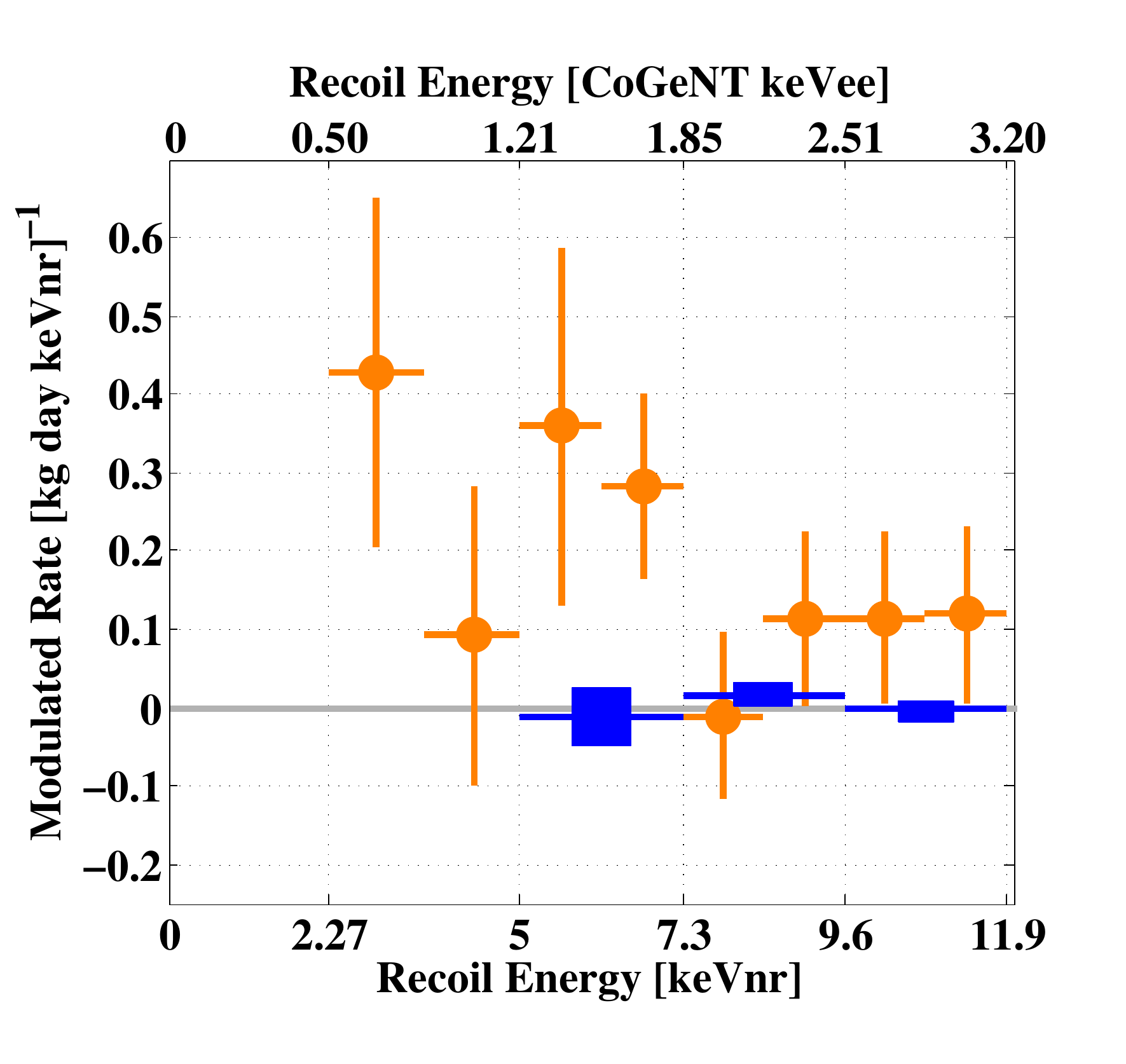}
\includegraphics[width=3.5in]{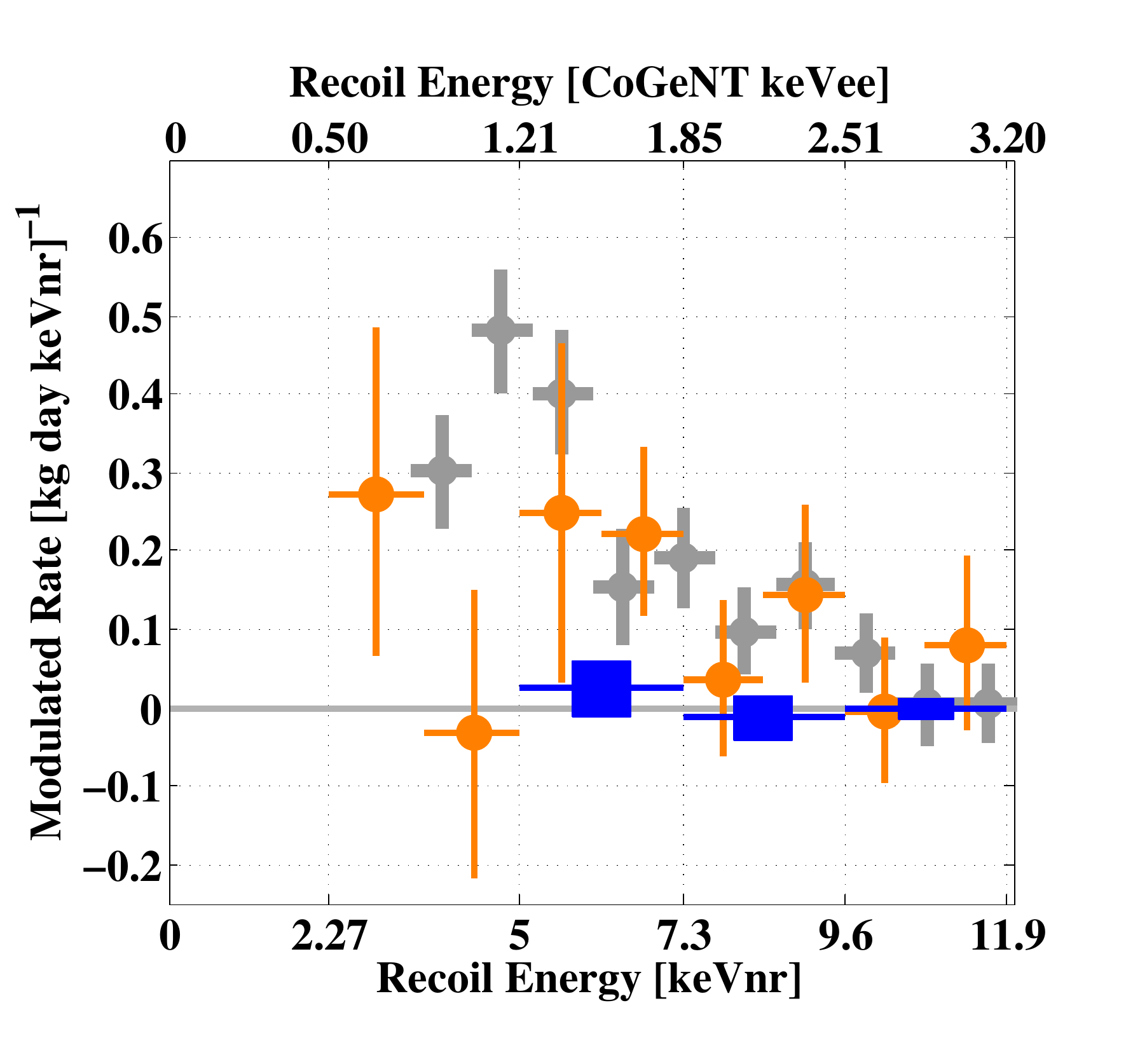}
\caption{\small Amplitude of modulation vs. energy, showing maximum-likelihood fits where the phase has been fixed and the modulated rates $M$ have been determined for both CoGeNT (light orange circles, vertical bars denoting the 68\% confidence intervals) and CDMS (dark blue rectangles, with vertical height denoting the 68\% confidence intervals).  The phase that best fits CoGeNT (106 days) over the full CoGeNT energy range is shown on the left; the phase expected from interactions with a generic WIMP halo (152.5 days) is shown on the right.  The upper horizontal scales show the electron-recoil-equivalent energy scale for CoGeNT events. The 5\textendash11.9\,keV$_{\mathrm{nr}}$ energy range over which this analysis overlaps with the low-energy channel of CoGeNT has been divided into 3 equal-sized bins (CDMS) and 6 equal-sized bins (CoGeNT).  In the right plot, we also show the DAMA modulation spectrum (small grey circles), following the method of Fox \emph{et al.}~\cite{Fox:2011integrateout}, for which we must assume both a WIMP mass (here, m${\chi}$=10~GeV/c$^{2}$) and a Na quenching factor (here, $q_{Na}=0.3$).  Lower WIMP masses or higher quenching factors can push the DAMA modulated spectrum towards significantly lower energies.  No attempt has been made to adjust for varying energy resolutions between the experiments.}
\label{fig:NR_rmodVenergy}
\end{center}
\end{figure}

\clearpage
\vspace{20mm}
\subsection{Rate Modulation:  Nuclear Recoil Singles}
\vspace{20mm}

\begin{figure}[htbp]
\begin{center}
\includegraphics[width=3.0in]{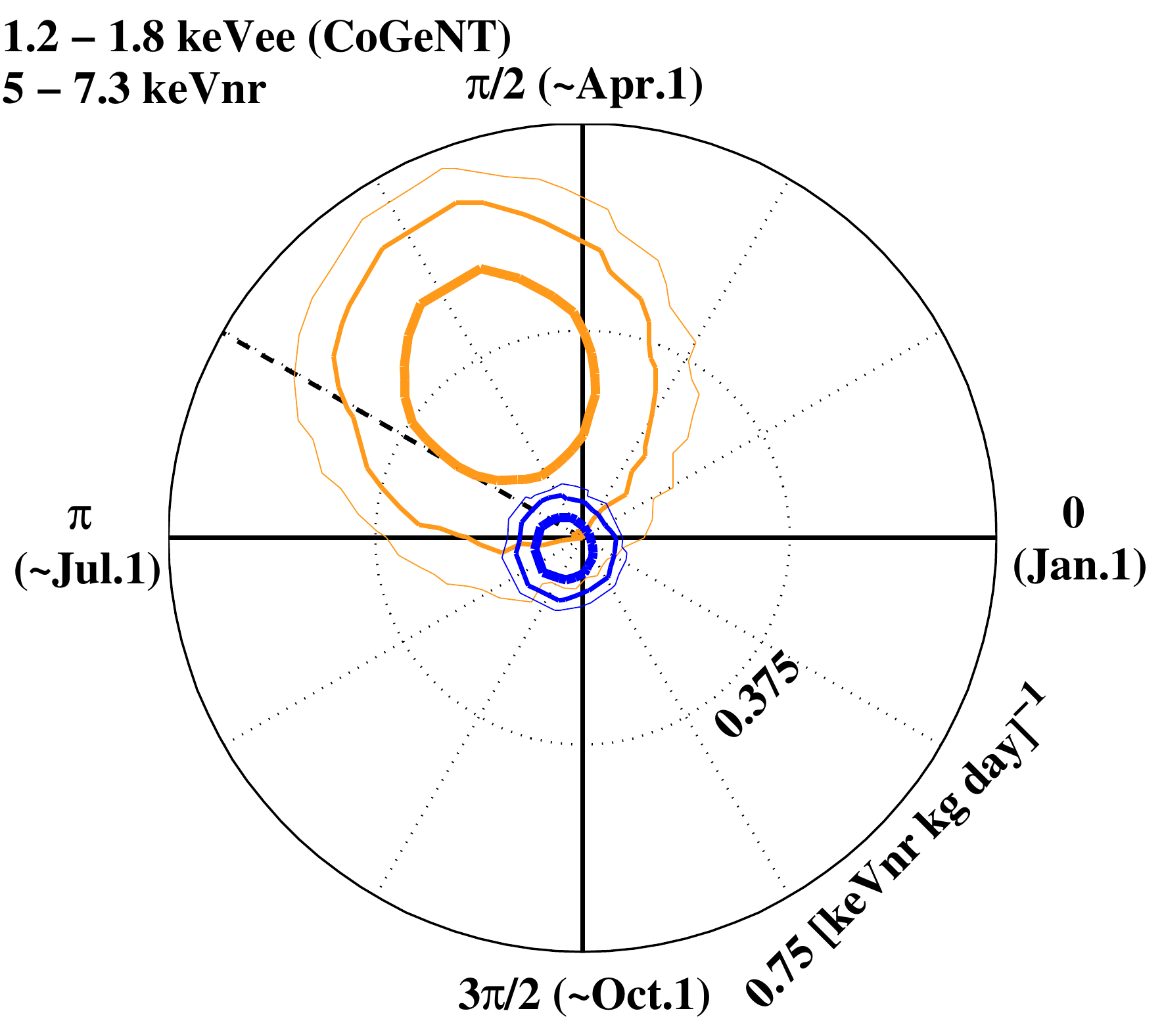}
\includegraphics[width=3.0in]{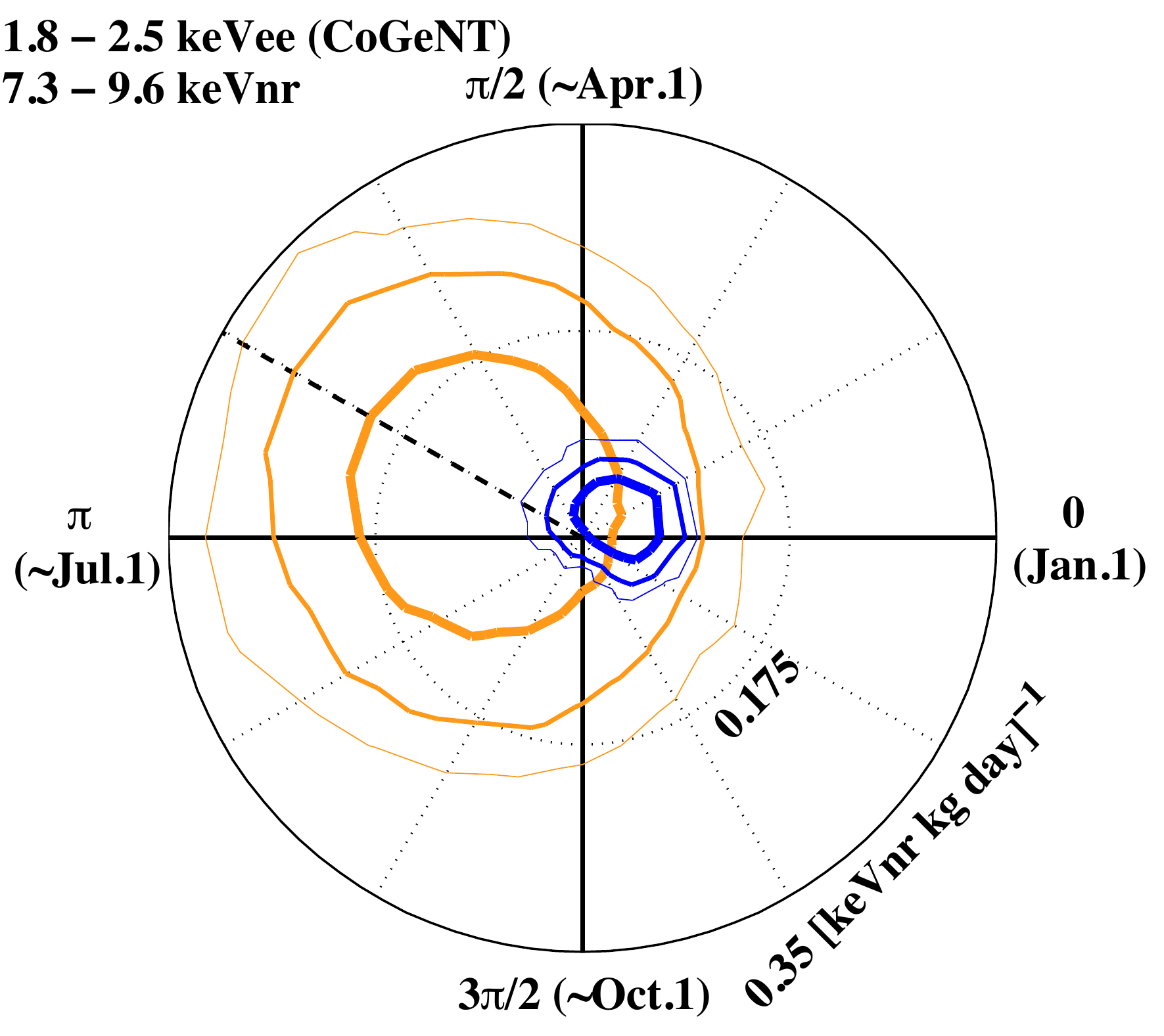}
\includegraphics[width=3.0in]{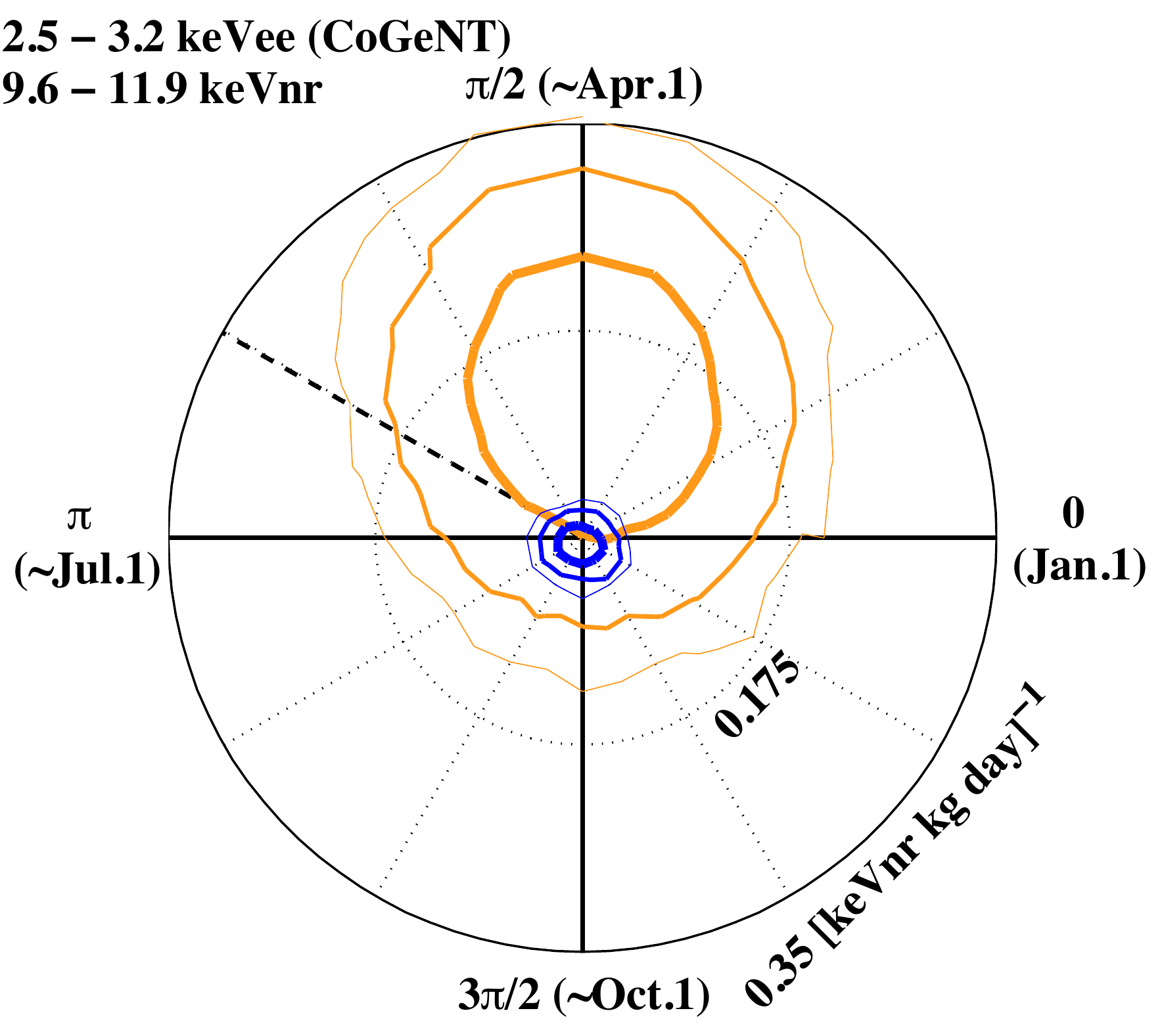}
\includegraphics[width=3.0in]{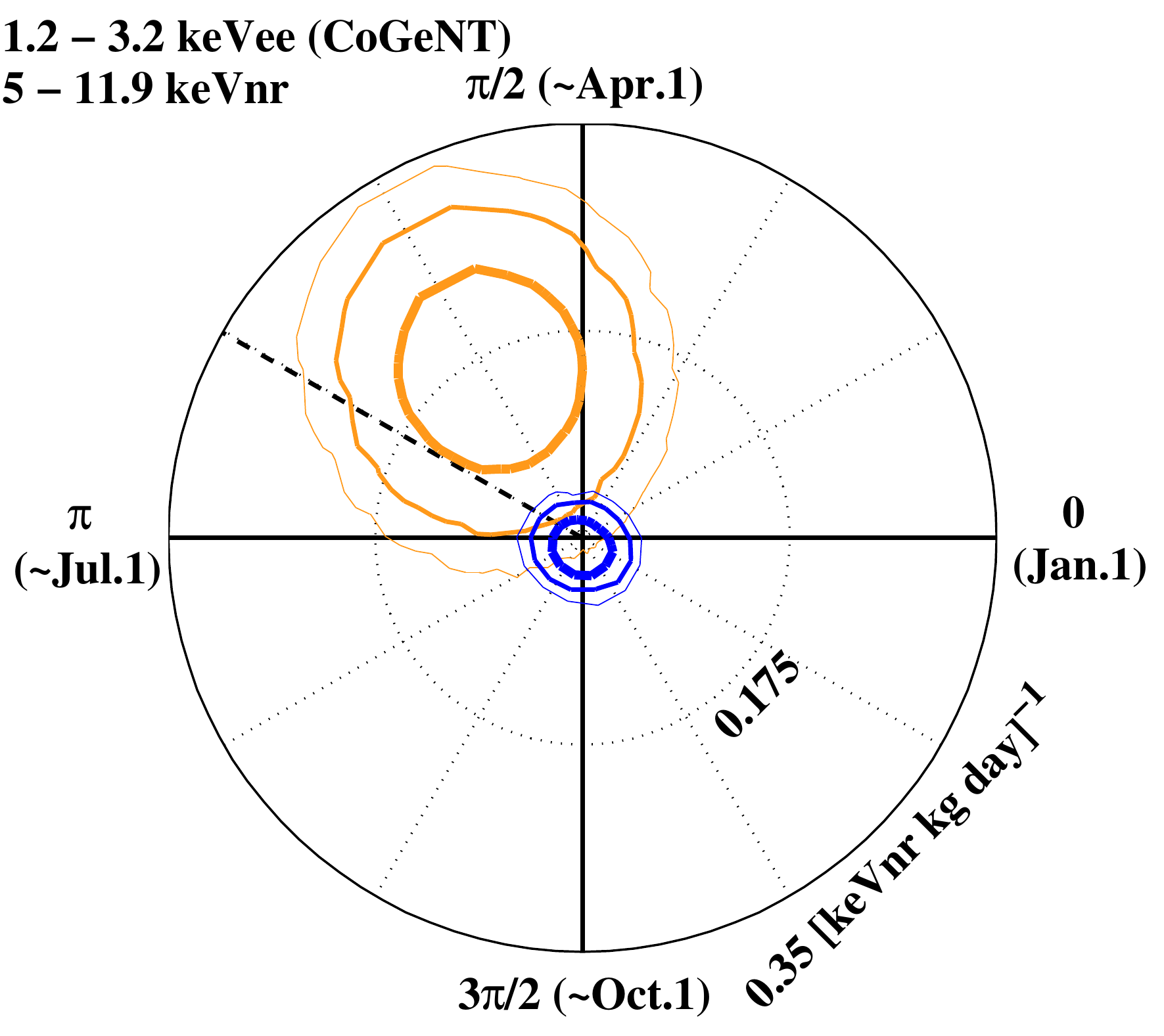}
\caption{\small 
Feldman-Cousins allowed regions in a polar projection of modulated rate $M$ vs. $\phi$ for CDMS singles passing the nuclear-recoil cut (dark blue) and for the CoGeNT data (light orange). Three different energy bins are shown, along with the total energy range (lower right).  Contours are 68, 95, and 99$\%$.}
\label{fig:NR_rmodVphase}
\end{center}
\end{figure}

\clearpage
\vspace{20mm}
\subsection{Rate Modulation:  Multiples and Singles}
\vspace{20mm}
\begin{figure}[htbp]
\begin{center}
\includegraphics[width=3.0in]{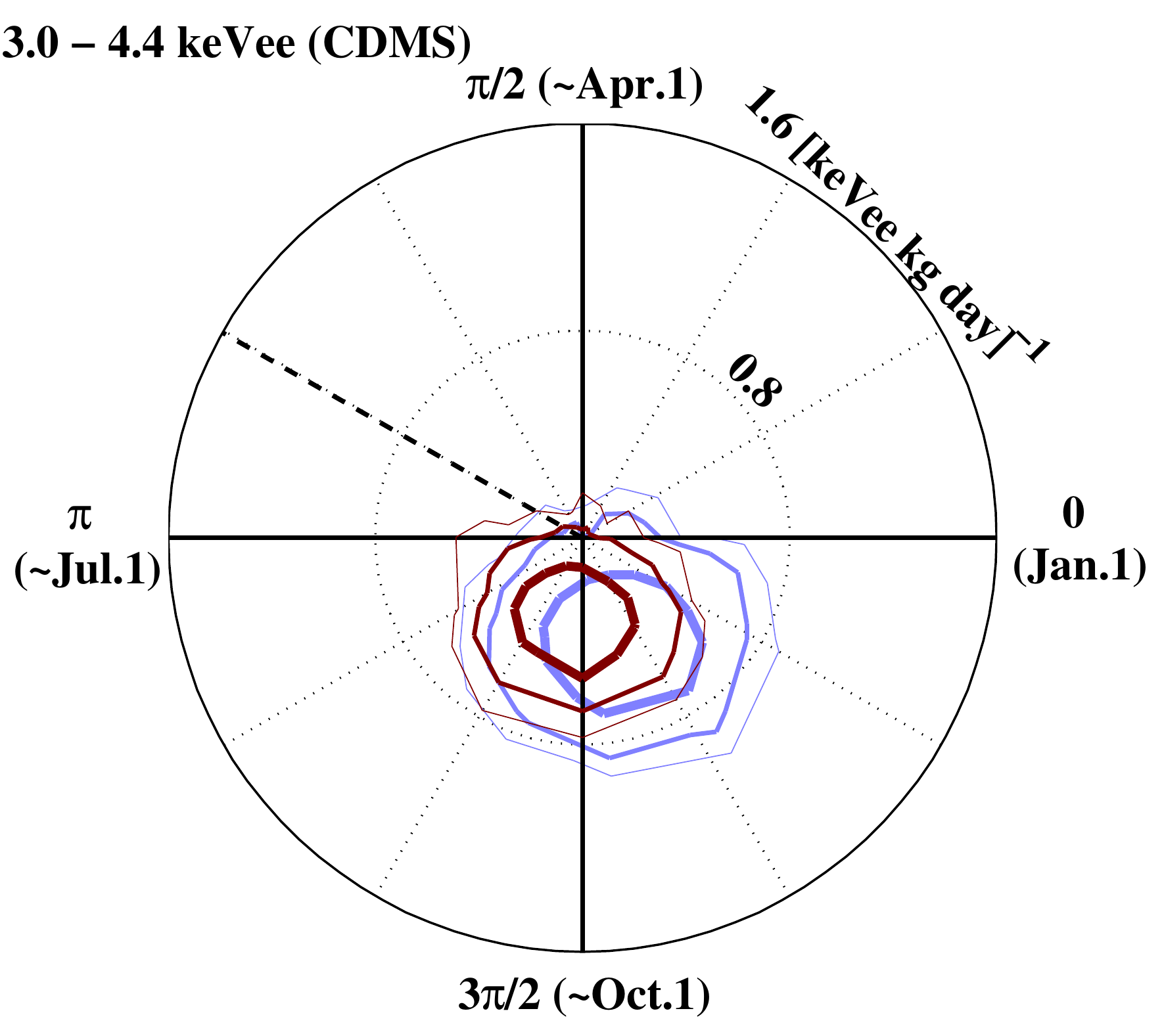}
\includegraphics[width=3.0in]{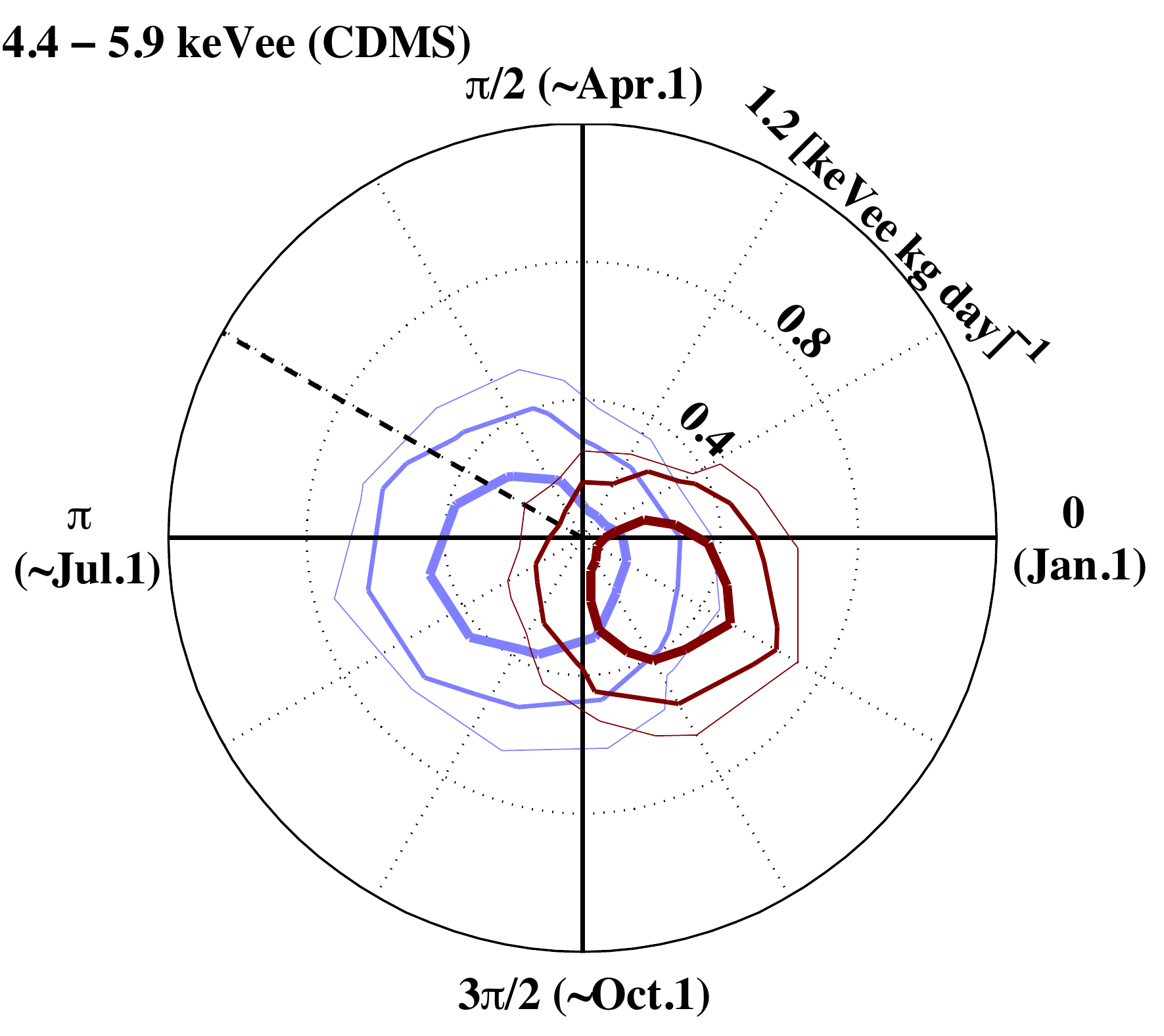}
\includegraphics[width=3.0in]{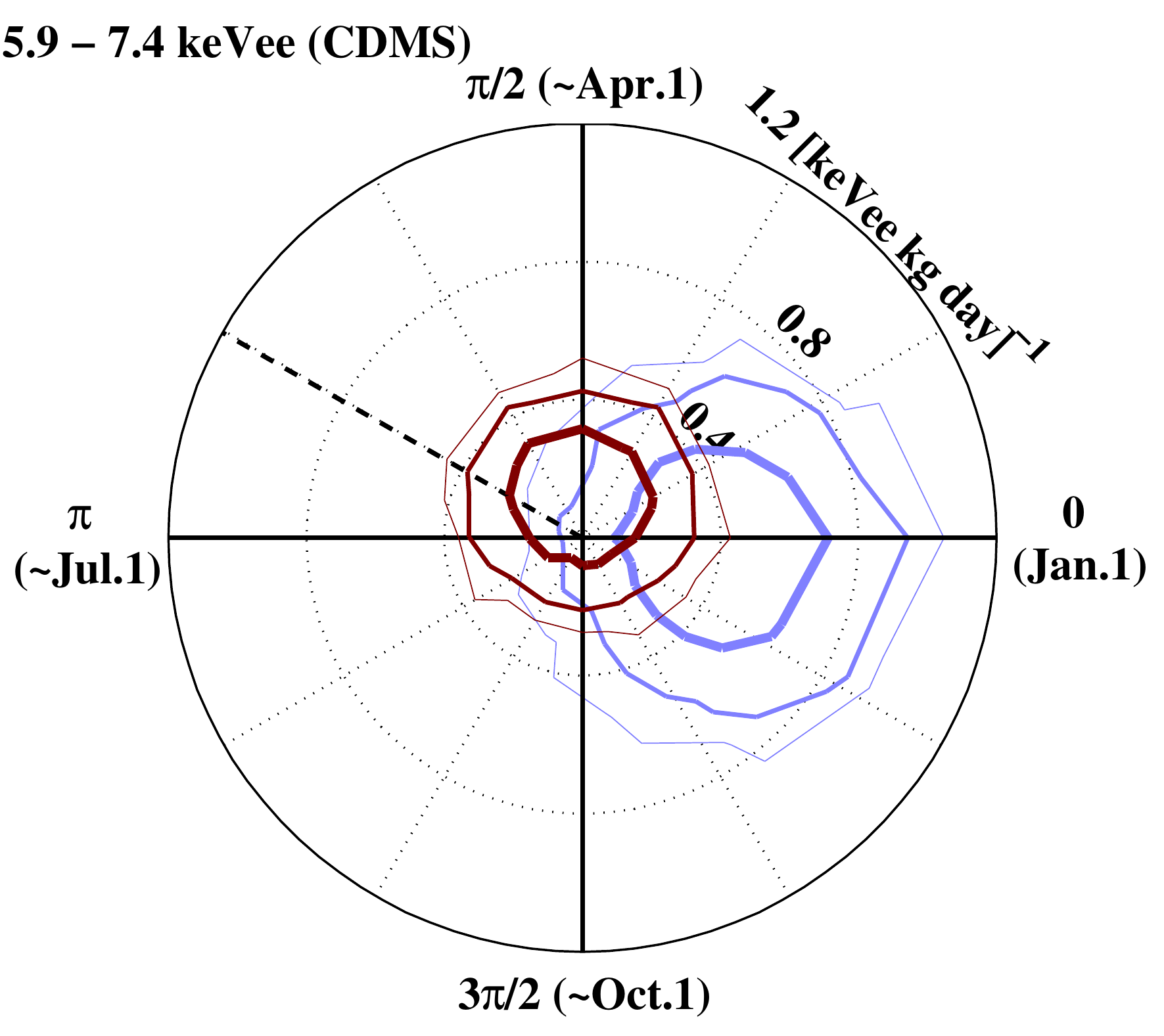}
\includegraphics[width=3.0in]{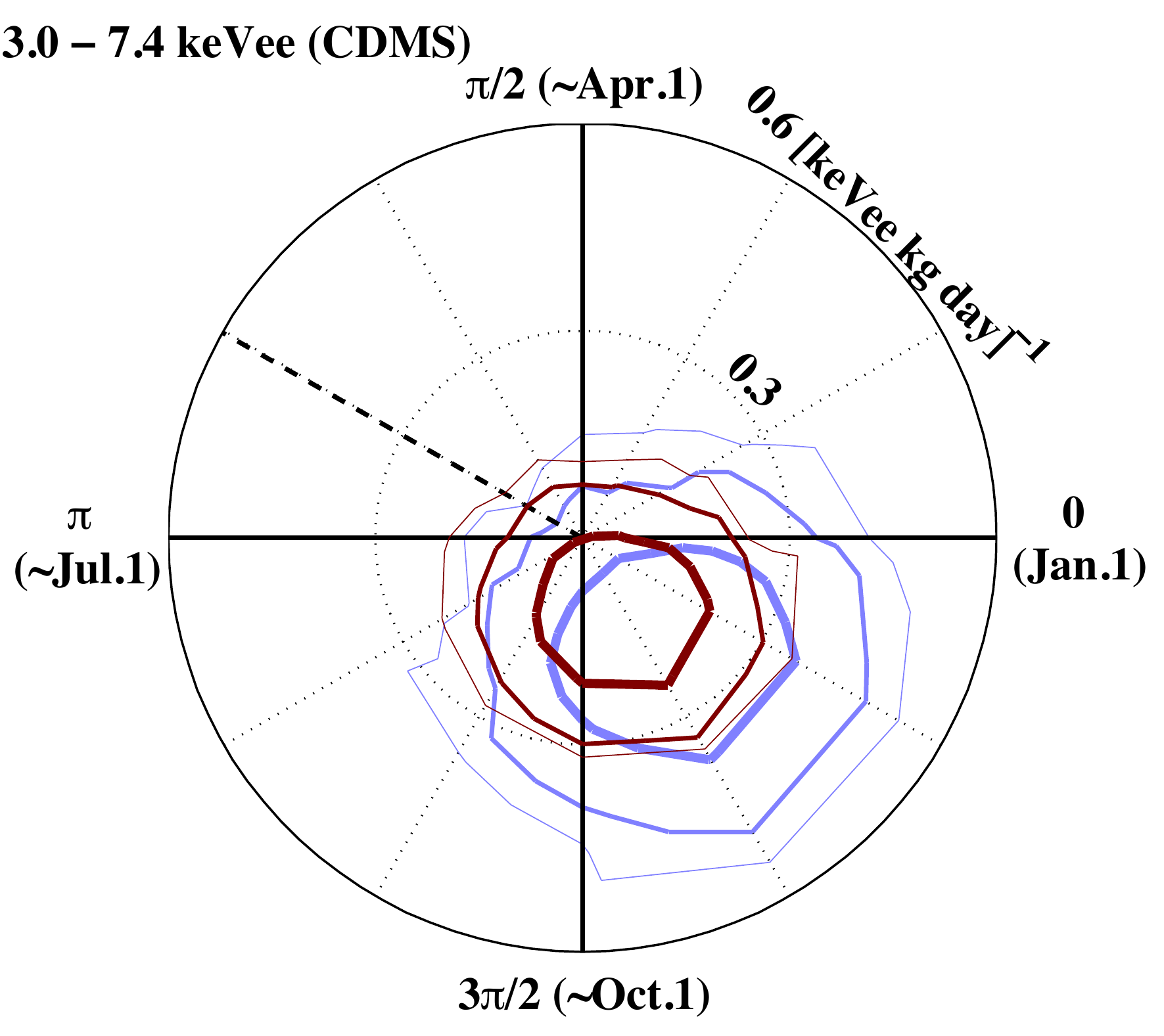}
\caption{\small 
Feldman-Cousins allowed regions in a polar projection of modulated rate $M$ vs. $\phi$ for two event populations dominated by electron recoils:    multiples (light blue) and singles (dark red). Three different energy bins are shown, along with the total energy range (lower right).  Contours are 68, 95, and 99$\%$.}
\label{fig:nonNR_rmodVphase}
\end{center}
\end{figure}

\end{document}